\def\R{\mathbb{R}}   

\newcommand{\e}{e^{-\rho \tau}}
\newcommand{\ee}{e^{-2 \rho \tau}}

\documentclass[12pt,a4paper]{article}
 
\usepackage{psfrag} 
\usepackage{graphicx}
\usepackage{amsmath}
\usepackage{exscale}
\usepackage{amsmath}
\usepackage{amsfonts}
\usepackage{amssymb}
\usepackage{latexsym}

\newtheorem{theorem}{Theorem}[section]
\newtheorem{proposition}[theorem]{Proposition}
\newtheorem{lemma}[theorem]{Lemma}
\newtheorem{corollary}[theorem]{Corollary}

\newtheorem{example}[theorem]{Example}
\newtheorem{remark}[theorem]{Remark}

\def\cF{\mathcal{F}}
\def\cC{\mathcal{C}}
 
\def\LaTeX{L\kern -.36em\raise .3ex\hbox{\sc a}\kern -.15em T\kern -.1667em%
\lower .7ex\hbox{E}\kern -.125em X}
 
\font\tt=cmtt10
 
\def\Proof{\bigskip\goodbreak\noindent{\it Proof: }}
\setbox1=\vbox{\hsize 5pt\vsize 5pt}
\def\boxitk#1{\vbox{\hrule\hbox{\vrule\kern4pt
          \vbox{\kern4pt#1\kern4pt}\kern4pt\vrule}\hrule}}
 
\def\sq{\hfill\boxitk{\box1}\goodbreak\bigskip}
 
\def\bbar#1{\overline{#1}}

\normalsize
\begin{document}
 
%\keywords{These are optional}
%\mathclass{Primary 46C20; Secondary 32G81.}
%\thanks{Research of the first author supported by KBN grant 00-000.}
%\abbrevauthors{A. Alfonsi ET AL.}
%\abbrevtitle{Optimal execution in LOB markets}
 
\title{Optimal execution strategies\\
in limit order books\\
with general shape functions}
 
\author{\small Aur\'elien Alfonsi\thanks{Supported  by
Deutsche Forschungsgemeinschaft through the Research Center {\sc
Matheon}
\lq\lq Mathematics for key technologies"  (FZT 86). }\\\small  CERMICS, projet
MATHFI\\\small   Ecole Nationale des Ponts et Chauss\'{e}es\\\small  6-8 avenue
Blaise Pascal\\\small  Cit\'{e} Descartes, Champs sur Marne\\\small  77455
Marne-la-vall\'{e}e, France\\\small  {\tt alfonsi@cermics.enpc.fr
}\setcounter{footnote}{6} 
\and\small Antje Fruth\\\small Quantitative
Products Laboratory\\\small Alexanderstr. 5\\\small 10178 Berlin,
Germany\\\small{\tt fruth@math.tu-berlin.de}
\and\small Alexander
Schied$^*$\\\small  Department of Mathematics, MA 7-4\\\small  TU Berlin\\\small
Strasse des 17. Juni 136\\\small  10623 Berlin, Germany\\\small  
{\tt schied@math.tu-berlin.de} 
}
 
\date{\small To appear in Quantitative Finance\\
Submitted September 3, 2007, accepted July 24, 2008\\ This version: November 20, 2009}
 
\maketitle
 
\vskip -0.7cm
{\noindent {\bf Abstract:} We consider optimal execution strategies
for block market orders placed in a limit order book (LOB). We build on the resilience
model proposed by Obizhaeva and Wang (2005) but allow for a  general shape of the LOB
defined via a given density function. Thus, we can allow for empirically observed
LOB shapes and obtain a nonlinear price impact of market orders.  We distinguish
two possibilities for modeling the resilience of the LOB after a large market order: the
exponential recovery of the number of limit orders, i.e., of the volume of the LOB, or the
exponential recovery of the bid-ask spread. We consider both of these resilience modes
and,  in each case, derive explicit optimal execution strategies in discrete time.
Applying our results to a block-shaped LOB, we obtain a new closed-form representation for
the optimal strategy of a risk-neutral investor, which explicitly solves the recursive scheme given in Obizhaeva and
Wang (2005). We also provide some evidence for the robustness of optimal strategies with
respect to the choice of the shape function and the resilience-type.}
 
%*******************************************************************************
\section{Introduction.}\label{Sec1} 
 
A common problem for   stock traders consists in unwinding large block orders of
shares, which can comprise up to twenty
percent of the daily traded volume of shares. Orders of this size create
significant impact on the asset price and, to reduce the overall market impact, it
is necessary to split them into smaller orders that are subsequently placed
throughout a certain time interval. The question at hand is thus to allocate an
optimal proportion of the entire order to each individual placement such that the
overall price impact is minimized.
 
Problems of this type were   investigated by Bertsimas and Lo~\cite{BertsimasLo},
Almgren and Chriss~\cite{AlmgrenChriss1,AlmgrenChriss2}, 
Almgren and Lorenz~\cite{AlmgrenLorenz}, Obizhaeva and Wang~\cite{ow}, and Schied and
Sch\"oneborn \cite{SchiedSchoeneborn,SchiedSchoeneborn2} to mention only a few. For extensions to  situations
with several competing traders, see~\cite{BrunnermeierPedersen},
\cite{CarlinLoboVishwanathan},
\cite{SchoenebornSchied}, and the references therein.
 
The mathematical formulation of the corresponding optimization problem relies
first of all on specifying a stock price model that takes into account the often
nonlinear feedback effects resulting from the placement of large orders by a \lq large
trader\rq. In the majority of models  in the literature, such orders affect the stock
price in the following two ways. A first part of the price impact is permanent and forever
pushes the price in a certain direction (upward for buy orders, downward for sell
orders). The second part, which is usually called the temporary impact, has no
duration and only instantaneously affects the trade that has triggered it. It is
therefore equivalent to a  (possibly nonlinear)
penalization by transaction costs. Models of this type underlie the
above-mentioned papers~\cite{BertsimasLo},  \cite{AlmgrenChriss1},  \cite{AlmgrenChriss2},
\cite{AlmgrenLorenz}, \cite{BrunnermeierPedersen}, \cite{CarlinLoboVishwanathan}, and~\cite{SchoenebornSchied}. Also the market impact models described in Bank and
Baum~\cite{BankBaum}, Cetin {\it et al.}~\cite{CetinJarrrowProtter}, Frey~\cite{Frey},
and Frey and Patie~\cite{FreyPatie} fall into that category.
While most of these models start with the dynamics of the asset price process as
a given fundamental, Obizhaeva and Wang~\cite{ow} recently proposed a market
impact model that derives its dynamics from an underlying model of a limit order book
(LOB). In
this model, the ask part of the LOB consists of a uniform distribution of shares offered
at prices higher than the current best ask price. When the large trader is not active, the
mid price of the LOB fluctuates according to the actions of noise traders, and the bid-ask
spread remains constant. A buy market order of the large trader, however,  consumes a
block of shares located immediately  to the right of the best ask and thus increase the
ask price by a linear proportion of the size of the order.   In addition, the LOB will
recover from the impact of the buy order, i.e., it will show a certain resilience. The
resulting price impact will neither be instantaneous nor entirely permanent but will decay
on an exponential scale. 
 
The model from~\cite{ow}  is quite close to descriptions of price impact on LOBs found in
empirical studies such as Biais \emph{et al.} \cite{Biais}, Potters and Bouchaud
\cite{PottersBouchaud}, Bouchaud \emph{et al.} \cite{Bouchaudetal}, and Weber and Rosenow
\cite{WeberRosenow}. In particular, the existence of a strong resilience effect, which
stems from the placement of new
limit orders close to the bid-ask spread, seems to be a well established fact, although
its quantitative features seem to be the subject of an ongoing discussion. 
 
In this paper, we will pick up the LOB-based market impact model from~\cite{ow} and
generalize it by allowing for a \emph{nonuniform} price distribution of shares within
the LOB. The resulting LOB shape which is nonconstant in the price conforms to empirical 
observations made in \cite{Biais, PottersBouchaud, Bouchaudetal, WeberRosenow}. 
It also leads completely naturally to a
\emph{nonlinear} price impact of market orders as found in an empirical
study by Almgren {\it et al.}~\cite{AlmgrenHauptman}; see also Almgren~\cite{ra}
and the references therein.  
In this generalized model, we will also consider the following two distinct possibilities
for modeling the  resilience of the LOB after a large market
order: the exponential recovery of
the number of limit orders, i.e., of the volume of the LOB (Model 1), or the exponential
recovery of the bid-ask spread (Model 2). While one can imagine also other possibilities,
we will focus on these two obvious resilience modes. Note that we assume the LOB shape to be
constant in time. Having a time-varying LOB shape will be an area of ongoing research.

We do not have a classical permanent price impact in our model for the following
reasons: Adding classical permanent impact, which is proportional to the volume traded,
 would be somewhat artificial in our model. In addition, this would not change optimal strategies as the optimization problem will be exactly the same as
without permanent impact. What one would want to have instead is a permanent
impact with a sensible meaning in the LOB context. But this would bring
  substantial difficulties in our derivation of optimal strategies.
 
After introducing the generalized LOB with its two resilience modes, we
consider the problem of optimally executing a buy order for $X_0$ shares within a
certain time frame
$[0,T]$. The focus on buy orders is for the simplicity of the presentation only,
completely analogous results hold for sell orders as well. While most other
papers, including~\cite{ow}, focus on optimization within the class of \emph{deterministic}
strategies, we will here allow for dynamic updating of trading strategies, that is,
we optimize over the larger class of
 {adapted} strategies. We will also allow for intermediate sell orders in our
strategies. Our main results, Theorem~\ref{prop1} and Theorem~\ref{prop3}, will
provide explicit solutions of this problem in Model 1 and Model 2, respectively.
Applying our results to a block-shaped LOB, we obtain a new closed-form
representation for the corresponding optimal strategy, which explicitly solves the
recursive scheme given in~\cite{ow}. Looking at several examples, we will also
find some evidence for the robustness of the optimal strategy. That is the optimal
strategies are qualitatively and quantitatively rather insensitive with respect to
the choice of the LOB shape. {In practice, this means that we can use them even
  though the LOB is not perfectly calibrated and has a small evolution during the
  execution strategy.}
 
The model we are using here is time homogeneous: the resilience rate is 
 constant and trading times are equally spaced.  By using the techniques introduced in our subsequent paper \cite{AFS}, it is possible to relax these assumptions and to allow for time inhomogeneities and also for linear constraints, at least in block-shaped models.

%Thus they might be constructed without knowing the shape
%of the LOB, which is empirically difficult to calibrate due to, for instance, the
%existence of hidden orders. 
 
The method we use in our proofs is  different from the approach used in~\cite{ow}.
Instead of using dynamic
programming techniques, we will first reduce the model of a full LOB with
nontrivial bid-ask spreads to a simplified model, for which the bid-ask spreads
have collapsed but the optimization problem is equivalent. The minimization of the
simplified cost functional is then reduced to the minimization of  certain
functions that are defined on an affine space. This latter minimization is then
carried out by means of the Lagrange multiplier method and explicit calculations.

The paper is organized as follows. In Section~\ref{Sec2}, we explain the two
market impact models that we derive from the generalized LOB model with different
resilience modes. In Section~\ref{costSection}, we set up the resulting
optimization problem. The main results for Models 1 and 2 are presented in the
respective Sections~\ref{Model1ResultsSection} and~\ref{Model2ResultsSection}. In
Section~\ref{Sec4}, we consider the special case of 
a uniform distribution of shares in the LOB as considered in~\cite{ow}. In
particular, we provide our new explicit formula for the optimal strategy in a
block-shaped LOB as obtained in~\cite{ow}. Section~\ref{Sec5} contains  numerical
and theoretical studies of the optimization problem for various nonconstant shape
functions. The proofs of our main results are given in the remaining
Sections~\ref{ReductionSection} through~\ref{App_OW}. More precisely, in Section
\ref{ReductionSection} we reduce the optimization problem for our two-sided LOB
models to the optimization over deterministic strategies within a simplified model
with a collapsed bid-ask spread. The derivations of the explicit forms of the
optimal strategies in Models 1 and 2 are carried out in the respective Sections~\ref{Model1proofsection} and~\ref{Model2proofsection}. In Section~\ref{App_OW} we prove the results for
block-shaped LOBs from Section~\ref{Sec4}.

\section{Two market impact models with resilience.}\label{Sec2}

In this section, we aim at  modeling the dynamics of a LOB that is
exposed to  repeated market orders by a large trader. The overall goal of the large trader will be to purchase a large amount
$X_0>0$ of shares within    a
certain time period
$[0,T]$. Hence, emphasis is on buy orders, and we  concentrate first on  the upper
part of the LOB, which consists of shares offered at various ask prices. The
lowest ask price at which shares are offered is called the
\emph{best ask price}. 
 
 Suppose first that the large trader is not active, so that
the dynamics of the limit order book are  determined by the
actions of noise traders only. We assume that  the corresponding unaffected
best ask price
$A^{{0}}$ is a martingale on a given
filtered probability space $(\Omega,(\cF_t),\cF,\mathbb{P})$ and satisfies
$A_0^{{0}}=A_0$. This assumption includes in particular the case in which $A^{{0}}$
is 
 a Bachelier model, i.e., 
$A_t^{{0}}=A_0+\sigma W_t$
for an $(\cF_t)$-Brownian motion $W$, as considered in~\cite{ow}. We emphasize, however,
that we can take any martingale and hence use, e.g.,  a geometric Brownian motion,
which avoids the  counterintuitive negative prices of the Bachelier model.
Moreover, we can allow for jumps in the dynamics of $A^{{0}}$ so as to model the
trading activities of other large traders in the market. In our context of a risk-neutral  investor minimizing the expected liquidation cost, the optimal strategies will turn out to be deterministic, due to the described martingale assumption.

Above the unaffected best ask price $A^0_t$, we assume a continuous
ask price distribution for available shares in the LOB: the number of shares
offered at price
$A_t^{{0}}+x$ is given by
$f(x)\,dx$ for a  continuous density function~$f:\R \longrightarrow ]0,\infty[$.
We will say that $f$ is the \emph{shape function} of the LOB. The choice of a
constant shape function
corresponds to the block-shaped LOB model of Obizhaeva and Wang~\cite{ow}.

\begin{figure}[htbp]
 \centering
 \includegraphics[width=0.7\linewidth]{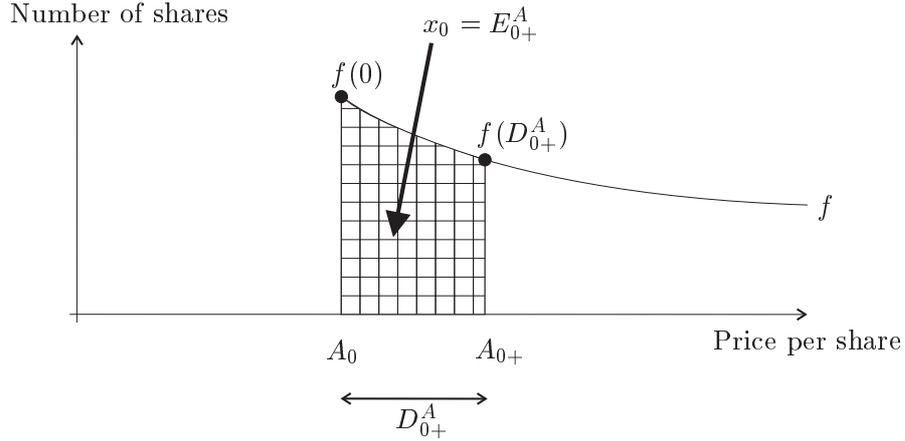} 
 \caption{The impact of a buy market order of $x_0$ shares .}
 \label{D and E}
\end{figure}

The shape function determines the impact of a market  order placed by our
large trader. Suppose for instance that the large trader places a buy market order
for
$x_0>0$ shares at time $t=0$. This market order will consume all shares located at
prices between $A_0$ and $A_0+D^A_{0+}$, where $D^A_{0+}$ is determined by 
$$\int_0^{D^A_{0+}} f(x)dx=x_0.
$$
Consequently, the ask price will be shifted up from $A_0$ to 
$$A_{0+}:=A_0+D^A_{0+};$$
see  Figure~\ref{D and E} for an illustration.
 
Let us denote by $A_t$ the actual ask price at time $t$, i.e., the ask
price after taking the price impact of previous buy orders  of the large trader
into account, and let us denote by 
$$D^A_t:=A_t-A_t^{{0}}
$$
the \emph{extra spread} caused by the actions of the large trader. Another buy market order of
$x_t>0$ shares will now consume all the shares offered at prices between $A_t$ and 
$$A_{t+}:=A_t+D^A_{t+}-D^A_{t}=A_t^{{0}}+D^A_{t+},
$$
where $D^A_{t+}$ is determined by the condition
\begin{equation}\label{Dt+definition}\int_{D^A_{t}}^{D^A_{t+}} f(x)dx=x_t.
\end{equation}
Thus, the process $D^A$  captures the impact of market orders on the current
best ask price. Clearly, the price impact $D^A_{t+}-D^A_{t}$ will be a nonlinear function of
the order size $x_t$ unless $f$ is constant between $D^A_t$ and $D^A_{t+}$. Hence, our model
includes the case of \emph{nonlinear impact functions}; see, e.g., Almgren~\cite{ra}
and Almgren {\it et al.}~\cite{AlmgrenHauptman} for a discussion.

Another important quantity is the process
\begin{equation}\label{D->E}
E^A_t=\int_{0}^{D^A_{t}}f(x)dx,
\end{equation}
of the number of shares
\lq already  eaten up\rq\ at time $t$. It quantifies the impact of the large
trader on the \emph{volume} of the LOB. By introducing the antiderivative 
\begin{equation}\label{(F)}F(z)=\int_0^z f(x)\,dx
\end{equation}
of $f$, the relation~(\ref{D->E}) can also be expressed as 
\begin{equation}\label{E->D}
E^A_t=F(D^A_t)\qquad\text{and}\qquad D^A_t=F^{-1}(E^A_t),
\end{equation}
where we have used our assumption that $f$ is strictly positive to obtain the
second identity. The relation~(\ref{Dt+definition}) is equivalent to
\begin{equation}\label{Et+definition}E^A_{t+} =E^A_t+x_t.
\end{equation}

 We still need to specify how $D^A$ and, equivalently, $E^A$ evolve when the
large trader is inactive in between market orders. 
It is a well established empirical fact that order books exhibit a certain resilience
as to the price impact of a large buy market orders, i.e., after the initial impact the
best ask price reverts back to its previous position; cf. Biais  \emph{et al.}
\cite{Biais}, Potters and Bouchaud
\cite{PottersBouchaud}, Bouchaud \emph{et al.} \cite{Bouchaudetal}, and Weber and Rosenow
\cite{WeberRosenow} for empirical studies.
That is, at least a part of the price impact will  only be temporary. For modeling this
resilience, we follow Obizhaeva and Wang~\cite{ow} in proposing an exponential recovery of
the LOB. While in the case of a block-shaped LOB as considered in~\cite{ow} the respective
assumptions of exponential recovery for 
$D^A$ and for
$E^A$ coincide, they provide two distinct possibilities for the case of a general
shape function. Since either of them appears to be plausible, we will discuss them
both  in the sequel. More precisely, we will consider the following two models for
the resilience of the market impact:
 
\medskip
 
\noindent {\bf Model 1:} The volume of the order book recovers 
 exponentially, i.e., $E$ evolves according to
\begin{equation}\label{Model1dynamics}E^A_{t+s}=e^{-\rho s}E^A_t
\end{equation}
if the large investor is inactive during the time interval $[t,t+s[$.
 
\medskip
 
\noindent {\bf Model 2:} The extra spread $D^A_t$ decays exponentially, i.e.,
\begin{equation}\label{Model2dynamics}D^A_{t+s}=e^{-\rho s}D^A_t
\end{equation} 
if the large investor is inactive during the time interval $[t,t+s[$.
 
\medskip
 
Here the \emph{resilience speed} $\rho$ is a positive constant, which for commonly
traded blue chip shares will often be calibrated such that the half-life time of the
exponential decay is in the order of a few minutes; see, e.g.,
\cite{PottersBouchaud, Bouchaudetal,  WeberRosenow}. Note that the dynamics of both
$D^A$ and $E^A$ are now completely specified in either model.
 
\medskip
 
Up to now, we have only described the effect of buy orders on the upper half of
the LOB. 
Since the overall goal of the larger trader is to buy $X_0>0$ shares up to time
$T$, a restriction to buy orders would seem to be reasonable. However,  we do not
wish to exclude the \emph{a priori} possibility that,  under
certain market conditions, it could be beneficial to also \emph{sell} some shares
and to buy them back at a later point in time. To this end, we also
need to model the impact of sell market orders on the lower part of the LOB, which
consists of a certain number of bids for shares at each price below the \emph{best
bid price}. As for ask prices, we will distinguish between an  unaffected
best bid price,
$B^0_t$, and the actual best bid price, $B_t$, for which  the price impact of
previous sell orders  of the large trader is taken into account. 
 All we assume on the dynamics of $B^0$ is
\begin{equation}\label{B0t}B_t^0\le A^0_t\qquad\text{at all times $t$.}
\end{equation}

The distribution of
bids below $B_t^0$ is modeled by the restriction of the shape function $f$ to the
domain
$]-\infty,0]$. More precisely,
for
$x<0$,  the number of bids at price
$B^0_t+x$ is equal to $f(x)\,dx$. The quantity
$$D^B_t:=B_t-B^0_t,
$$
 which usually will be negative, is called the \emph{extra spread} in the bid
price distribution.  A sell market order of
$x_t<0$ shares placed at time $t$ will  consume all the shares offered at prices
between
$B_t$ and 
$$B_{t+}:=B_t+D^B_{t+}-D^B_{t}=B_t^{{0}}+D^B_{t+},
$$
where $D^B_{t+}$ is determined by the condition
\begin{equation}\label{DBt+definition}x_t=\int_{D^B_{t}}^{D^B_{t+}}
f(x)dx=F(D^B_{t+})-F(D^B_{t})=E^B_{t+}-E^B_t,
\end{equation}
for $E^B_s:=F(D^B_s)$. Note that  $F$ is
defined via \eqref{(F)} also for negative arguments. If the large trader is
inactive during the time interval
$[t,t+s[$, then the processes $D^B$ and $E^B$ behave just as their counterparts
$D^A$ and
$E^A$, i.e.,
\begin{equation}\label{Bdynamics}\begin{split}E^B_{t+s}&=e^{-\rho
s}E^B_t \qquad\text{in Model 1,}\\
D^B_{t+s}&=e^{-\rho s}D^B_t \qquad\text{in Model 2.}
\end{split}
\end{equation}
 
\medskip

\section{The cost minimization problem.}\label{costSection}
 
When placing a single buy market order of size $x_t\ge0$ at time $t$, the large
trader will purchase
$f(x)\,dx$ shares at price $A^0_t+x$, with $x$ ranging from $D^A_t$ to
$D^A_{t+}$. Hence, the total cost of the buy market order amounts to 
\begin{eqnarray}\label{pit}
\pi_t(x_t):=\int_{D^A_t}^{D^A_{t+}}(A_t^{{0}}+x)f(x)\,dx=
A^{{0}}_tx_t+\int_{D^A_t}^{D^A_{t+}}xf(x)\,dx.
\end{eqnarray}
For a sell market order $x_t\le0$, we have
\begin{eqnarray}\pi_t(x_t):=B^{{0}}_tx_t+\int_{D^B_t}^{D^B_{t+}}xf(x)\,dx.
\label{Bpit}
\end{eqnarray}
 
 In practice, very large orders are often split into a number of
consecutive market orders to reduce the overall price impact. Hence, the question at hand
is to determine the size of the individual orders so as to minimize a cost criterion. So
let us assume that the large trader needs to buy a total of $X_0>0$ shares until time $T$
and that trading can occur at
$N+1$ equidistant times~$t_n=n \tau$ for~$n=0,\dots,N$ and~$\tau:={T}/{N}$. An
\emph{admissible strategy} will be a sequence $\xi=(\xi_0,\xi_1,\dots,\xi_N)$ of random
variables such that
\begin{itemize}
\item $\sum_{n=0}^N\xi_n=X_0$, 
\item each $\xi_n$ is measurable with respect to $\cF_{t_n}$,
\item each $\xi_n$ is bounded from below.
\end{itemize}
 The quantity $\xi_n$ corresponds to the
size of the market order placed at time $t_n$. Note that we do not \emph{a priori} require
$\xi_n$ to be positive, i.e., we also allow for intermediate sell orders, but we assume
that there is some lower bound on sell orders. 
 
The \emph{average cost} $\cC(\xi)$ of an admissible  strategy $\xi$ is defined as the
expected value of the total costs incurred by the consecutive market orders:
\begin{equation}\label{costfunctional}
\cC(\xi)=\mathbb{E}\Big[\,\sum_{n=0}^N\pi_{t_n}(\xi_n)\,\Big].
\end{equation}
 Our
goal in this paper consists in finding admissible strategies that minimize the average cost
within the class of all admissible strategies. For the clarity of the exposition, we decided no to treat the case of a risk averse investor. We suppose that the introduction of risk aversion will have a similar effect as in \cite{ow}.

Note that the value of $\cC(\xi)$ depends on
whether we choose Model 1 or Model~2, and it will turn out that also the
quantitative---though not the qualitative---features of the optimal strategies will be
slightly model-dependent.
 
Before turning to the statements of our results, let us introduce the following standing
assumption for our further analysis:
 the function~$F$ is  supposed to be unbounded in the sense that
\begin{equation}\label{F unbounded}
\lim_{x \uparrow \infty}F(x)=\infty \hspace{0.3 cm} \mbox{and} \hspace{0.3 cm} \lim_{x
\downarrow -\infty}F(x)=-\infty.
\end{equation}
This assumption of unlimited order book depth is of course an idealization of reality and
is for convenience only. It should not make a difference, however, as soon as the depth of
the real LOB is big enough to accommodate every market order of our optimal strategy.

\section{Main theorem for Model 1.}\label{Model1ResultsSection}
We will now consider the minimization of the cost functional $\cC(\xi)$ in  Model 1, in
which we assume an exponential recovery of the LOB volume; cf.~(\ref{Model1dynamics}).
 
\medskip
 
\begin{theorem}\label{prop1}{\rm (Optimal strategy in Model 1).\\}
Suppose that the function $h_1: \R \rightarrow \R_+$ with
\[h_1(y):=F^{-1}(y)-\e F^{-1}(\e y) \]
is one-to-one. Then there exists a unique optimal strategy $\xi^{(1)}=(\xi_0^{(1)},\dots,\xi^{(1)}_N)$.
The initial market order $\xi_0^{(1)}$ is the unique solution of the equation
\begin{equation}\label{x_0 condition 1}
F^{-1}\left(X_0-N \xi^{(1)}_0\left(1-\e\right)\right)=\frac{h_1(\xi^{(1)}_0)}{1-\e},
\end{equation}
the intermediate orders are given by
 \begin{equation}\label{strategy}
\xi^{(1)}_1=\dots=\xi^{(1)}_{N-1}=\xi^{(1)}_0\left(1-\e\right), 
\end{equation}
and the final order is determined by
$$ \xi^{(1)}_N=X_0-\xi^{(1)}_0-(N-1)\xi^{(1)}_0\left(1-\e\right).
$$
In particular, the optimal strategy is deterministic. Moreover, it consists only of
nontrivial buy orders, i.e., $\xi^{(1)}_n>0$ for all $n$.
\end{theorem}

\medskip
 
Some remarks on this result are in order. First, the optimal strategy $\xi^{(1)}$
consists only of buy orders and so the bid price remains unaffected, i.e., we have
$E^B_t\equiv0\equiv D^B_t$. It follows moreover that the process $E:=E^A$ is
recursively given by the following Model 1 dynamics:
\begin{eqnarray}
E_{0}&=&0,\nonumber\\
 E_{t_n+}&=&E_{t_n}+\xi^{(1)}_n, \qquad n=0,\dots,N,\label{E}\\
E_{t_{k+1}}&=&\e E_{t_k+} =\e  (E_{t_k}+\xi^{(1)}_k ),\qquad
k=0,\dots,N-1.\nonumber
\end{eqnarray}
Hence, by \eqref{x_0 condition 1} and \eqref{strategy},
\begin{equation}\label{E2}
E_{t_n+}=\xi^{(1)}_0 \hspace{0.3 cm}\text{and}\hspace{0.3 cm}E_{t_{n+1}}=\e\xi^{(1)}_0 
\hspace{0.3 cm} \mbox{for
$n=0,\dots,N-1$.}
\end{equation}
That is, once $\xi^{(1)}_0$ has been determined via~(\ref{x_0 condition 1}), the optimal
strategy consists in a sequence of market orders that consume exactly that amount of
shares by which  the LOB has recovered since the preceding market order, due to the
resilience effect. At the terminal time $t_N=T$, all remaining shares are bought. In the
case of a block-shaped LOB, this qualitative pattern was already observed by Obizhaeva and
Wang~\cite{ow}. Our Theorem~\ref{prop1} now shows that this optimality pattern is actually independent of the LOB
shape, thus indicating a certain robustness of  optimal strategies.

\medskip

\begin{remark}{
According to \eqref{E->D} and \eqref{E2}, the extra spread $D:=D^A$  of the optimal
strategy
$\xi^{(1)}$ satisfies
$$D_{t_n+}=F^{-1}\left(E_{t_n+}\right)=F^{-1}(\xi^{(1)}_0).
$$
For $n=N$ we moreover have that
\begin{eqnarray*}
D_{t_N+}&=&F^{-1}\left(E_{t_N+}\right)=F^{-1}\left(E_{t_N}+\xi^{(1)}_N\right)\\
&=&F^{-1}\left(\xi^{(1)}_0 \e+X_0-\xi^{(1)}_0-(N-1)\xi^{(1)}_0 \left(1-\e\right)\right)\\
&=&F^{-1}\left(X_0-N \xi^{(1)}_0\left(1-\e\right)\right).
\end{eqnarray*}
Hence, the left-hand side of \eqref{x_0
condition 1} is equal to~$D_{t_N+}$.
}\end{remark}

\medskip
 
We now comment on the conditions in Theorem~\ref{prop1}.

\medskip

\begin{remark}\label{rem1}{\rm (When is~$h_1$ one-to-one?) }{The function $h_1$ is
continuous with~$h_1(0)=0$ and~$h_1(y)>0$ for~$y>0$. Hence,~$h_1$ is one-to-one
if and only if~$h_1$ is strictly increasing. We want to consider when this is the case. To
this end, note that the condition
\[h_1'(y)=\frac{1}{f(F^{-1}(y))}-\frac{\ee}{f(F^{-1}( \e y))}>0\]
 is equivalent to 
\begin{equation}\label{inequ}
\ell(y):=f(F^{-1}(\e y))-\ee f(F^{-1}(y))>0.
\end{equation}
 That is, the function~$h_1$ will be one-to-one if, for instance, the  shape
function~$f$ is decreasing for~$y> 0$ and increasing for~$y <0$. 
In fact, it has been observed in the empirical studies
\cite{Biais, PottersBouchaud,
Bouchaudetal,  WeberRosenow} that average shapes of typical order books have a maximum at
or close to the best quotes and then decay as a function of the distance to the best
quotes, which would conform to our assumption.

%% \begin{figure}[htbp]
%%  \centering
%%  \includegraphics[width=0.6\linewidth]{Figure2b.eps}
 %\scalebox{0.7}{\includegraphics{Grafik3.eps}}  
%%  \caption{A shape function~$f$ for which the corresponding
%% function~$h_1$ is not one-to-one.}
%%  \label{f}
%% \end{figure}
 
%% We now want to give an example of a shape function~$f$ such that the corresponding
%% function~$h_1$ is not one-to-one. To this end, let us assume for simplicity that there
%% exists an~$n\in
%% \{2,3,\dots\}$ such that~$\e=\frac{1}{n}$. We set  
%% \[f(x):= \begin{cases} 1 &  x\in [0,\frac{\frac{1}{2}n^2+1}{n-1}) \\
%% 1+n^2(x-\frac{\frac{1}{2}n^2+1}{n-1}) & 
%% x\in \left[\frac{\frac{1}{2}n^2+1}{n-1},\frac{\frac{1}{2}n^2+1}{n-1}+1 \right] \\
%% n^2+1 & x\in
%% \left(\frac{\frac{1}{2}n^2+1}{n-1}+1,\infty\right); \end{cases}\]
%% see Figure~\ref{f}. For 
%% $u:=(\frac{\frac{1}{2}n^2+1}{n-1}+1)+\frac{1}{2}n^2$
%% we then have $F^{-1}(u)=\frac{\frac{1}{2}n^2+1}{n-1}+1$ as well as 
%% \[F^{-1}(\e u)=F^{-1}\Big(\frac{u}{n}\Big)=\frac{\frac{1}{2}n^2+1}{n-1}.\]
%% Hence, we get
%% \[f\left(F^{-1}\left(\e u\right)\right)=1<\Big(1+\frac{1}{n^2}\Big)=\ee
%% f\left(F^{-1}(u)\right),\]
%%     which tells us according to~(\ref{inequ}) that the corresponding function~$h_1$ is not
%%     strictly increasing.
}
\end{remark}

\begin{remark}{\rm (Continuous-time limit of the optimal
strategy).}
\label{rem_lim_model1} One can also investigate the asymptotic behavior
of the optimal strategy  when the number $N$ of trades in~$]0,T]$ tends to
infinity. 
 It is not difficult  to see that $h_1/(1-\e)$ converges pointwise 
  to
$$h_1^{\infty}(y):=F^{-1}(y)+\frac{y}{f(F^{-1}(y))}.$$
 Observe also that
  $N(1-\e)\rightarrow \rho T$. Since for any~$N$ we have~$\xi^{(1)}_0 \in
]0,X_0[$,
  we can extract a subsequence that converges and its limit is then necessarily solution
of the equation
  $$ F^{-1}(X_0-\rho T y)=h_1^{\infty}(y).$$
  If this equation has a unique solution~$\xi^{(1),\infty}_0$ we deduce that the optimal initial
  trade converges to~$\xi^{(1),\infty}_0$ when $N\longrightarrow   \infty$. This is
the
  case, for example, if $h_1^{\infty}$ is strictly increasing and especially when
$f$
  is decreasing. In that case,
  $N\xi^{(1)}_1$ converges to $\rho T \xi^{(1),\infty}_0$ and $\xi^{(1)}_N$
  to $\xi^{(1),\infty}_T:=X_0-\xi^{(1),\infty}_0(1+\rho T)$. Thus, in
  the continuous-time limit, the optimal
strategy consists in an initial block order of $\xi^{(1),\infty}_0$ shares at
time~$0$, continuous buying at the constant rate $\rho \xi^{(1),\infty}_0 $
during $]0,T[$, and a final block order of 
  $\xi^{(1),\infty}_T$ shares at time~$T$.
\end{remark}  
 
\section{Main theorem for Model 2.}\label{Model2ResultsSection}
We will now consider the minimization of the cost functional 
$$\cC(\xi)=\mathbb{E}\Big[\,\sum_{n=0}^N\pi_{t_n}(\xi_n)\,\Big]
$$ 
in  Model 2, where we assume an exponential recovery of the extra spread; cf.
(\ref{Model2dynamics}).

\medskip

\begin{theorem}\label{prop3}{\rm (Optimal strategy in Model 2).\\}{ Suppose that the
function $h_2:\R \rightarrow \R$ with
\[h_2(x):=x \frac{f(x)-\ee f(\e x)}{f(x)-\e f(\e x)} \]
is one-to-one and that the shape function satisfies
\begin{equation}\label{explosion assumption}
\lim_{|x|\rightarrow \infty}  x^2 \inf_{z \in [\e x,x]}f(z) =\infty .
\end{equation}
Then there exists a unique optimal strategy $\xi^{(2)}=(\xi_0^{(2)},\dots,\xi^{(2)}_N)$.
The initial market order $\xi_0^{(2)}$ is the unique solution of the equation
\begin{equation}\label{x_0 condition 2}
F^{-1}\left(X_0-N
\big[\xi^{(2)}_0-F\big(\e
F^{-1}(\xi^{(2)}_0)\big)\big]\right)=h_2\big(F^{-1}(\xi^{(2)}_0)\big),
\end{equation}
the intermediate orders are given by
\begin{equation}\label{strat}
\xi^{(2)}_1=\dots=\xi^{(2)}_{N-1}=\xi^{(2)}_0 -F\big(\e F^{-1}(\xi^{(2)}_0)\big),
\end{equation}
and the final order is determined by
$$\xi^{(2)}_N=X_0-N \xi^{(2)}_0+(N-1)F\big(\e F^{-1}(\xi^{(2)}_0)\big).
$$
In particular, the optimal strategy is deterministic. Moreover, it  consists only of
nontrivial buy orders, i.e., $\xi^{(2)}_n>0$ for all $n$.
 }
\end{theorem}

\medskip
 
Since the optimal strategy $\xi^{(2)}$ consists only of buy orders, the processes
$D^B$ and $E^B$ vanish, and $D:=D^A$ is given by
\begin{eqnarray}
D_{0}&=&0,\nonumber\\ 
D_{t_n+}&=&F^{-1}\left(\xi^{(2)}_n+F\left(D_{t_n}\right)\right),\qquad
n=0,\dots,N\label{delta dynamic}\\
 D_{t_{k+1}}&=&\e D_{t_k+},\qquad k=0,\dots, N-1.\nonumber
\end{eqnarray}
Hence, induction shows that
\[D_{t_n+}= F^{-1}(\xi^{(2)}_0) \hspace{0.5 cm}\text{and} \hspace{0.5 cm}D_{t_{n+1}}=\e
F^{-1}(\xi^{(2)}_0) \hspace{0.5 cm}
\mbox{for~$n=0,\dots,N-1$.}\]
By~(\ref{E->D}), the process $E:=E^A$ satisfies
$$E_{t_n+}= \xi^{(2)}_0\hspace{0.5 cm}\text{and} \hspace{0.5 cm}E_{t_{n+1}}=F\big(\e
F^{-1}(\xi^{(2)}_0)\big) \hspace{0.5 cm}\mbox{for~$n=0,\dots,N-1$.}
$$ 
This is very similar to our result~(\ref{E2}) in Model 1: once $\xi^{(1)}_0$ has been
determined via~(\ref{x_0 condition 1}),  
the optimal strategy consists in a sequence of market orders that consume exactly that
amount of shares by which  the LOB has recovered since the preceding market order. At the
terminal time $t_N=T$, all remaining shares are bought. The only differences are in the
size of the initial market order and in the mode of recovery. This qualitative similarity
between the optimal strategies in Models 1 and 2 again confirms our observation of the
robustness of the optimal strategy.
 
\medskip
 
\begin{remark}{At the terminal time $t_N=T$, the extra spread is given by
\begin{eqnarray*}
D_{t_N+}&=&F^{-1}\left(E_{t_N+}\right)=F^{-1}\big(E_{t_N}+\xi^{(2)}_N\big)\\
&=&F^{-1}\left(X_0-N
\big[\xi^{(2)}_0-F\big(\e
F^{-1}(\xi^{(2)}_0)\big)\big]\right),
\end{eqnarray*}
and this expression coincides with the left-hand side in~(\ref{x_0 condition 2}).
}
\end{remark}
 
\medskip

Let us now comment on the conditions assumed in Theorem~\ref{prop3}. To this end, we first
introduce the function
\begin{equation}\label{widetildeF}
\widetilde{F}(z):=\int_0^z x f(x)dx.
\end{equation}
 
\medskip
 
\begin{remark}{If $\widetilde{F}$ is convex then condition~(\ref{explosion assumption})
in Theorem~\ref{prop3} is satisfied. This fact admits the following short proof. Take  $x^*
\in [\e x, x]$ realizing the infimum of $f$ in $[\e
x, x]$. Then 
\begin{equation}\label{ineq}
x^2 \inf_{z \in [\e x,x]}f(z)=x^2 f(x^*) \geq x^* \left( x^* f(x^*)\right).
\end{equation}
Due to the convexity of $\widetilde{F}$, its derivative $\widetilde{F}'(x)=x
f(x)$ is increasing. It is also  nonzero iff $x\neq0$. Therefore the right-hand side
of~(\ref{ineq}) tends to infinity for 
$|x|\rightarrow \infty$.}
\end{remark}
 
\medskip

However, the convexity of $\widetilde{F}$ is not necessary for condition~(\ref{explosion
assumption}) as is illustrated by the following simple example.
 
\medskip
 
\begin{example} 
Let us construct a shape function for which~(\ref{explosion assumption}) is satisfied
even though 
$\widetilde{F}$ need not be convex. To this end, take any   continuous function $b:\R
\rightarrow ]0,\infty[$ that is bounded away from zero. Then let
\[f(x):= \begin{cases} b(1)&  |x|\leq 1 \\
\frac{b(x)}{\sqrt{|x|}}&  |x|>1. \end{cases}\]
This shape function clearly satisfies condition~(\ref{explosion assumption}).
Taking for example
$b(x)=1+\varepsilon\cos(x)$ with $0<\varepsilon<1$, however, gives a nonconvex function
$\widetilde{F}$. Moreover, by choosing $\varepsilon$ small enough, we can obtain
$h'_2(x)>0$ so that the shape function $f$ satisfies the assumptions of Theorem~\ref{prop3}.
\end{example}

\medskip
 
We now comment on the condition that $h_2$ is one-to-one. The following example shows that
this is indeed a nontrivial assumption.

\medskip

\begin{example}{
We now provide an example of a  shape function~$f$ for which the corresponding
function~$h_2$ is not one-to-one. First note that  $h_2(0)=0$ and 
\begin{equation}\label{h2increasing}\lim_{\epsilon \downarrow 0}
    \frac{h_2(\epsilon)-h_2(0)}{\epsilon}=\frac{1-\ee}{1-\e}>0.
\end{equation}
Therefore and since~$h_2$ is continuous, it cannot be one-to-one if we can 
find~$x^*>0$ such that~$h_2(x^*)<0$. To this end, we assume
that there exist~$n\in
\{2,3,\dots\}$ such that~$\e=\frac{1}{n}$ and take 
\[f(x):= \begin{cases} (n+1)&  x\in \left[0,\frac{1}{n}\right) \\
(n+1)-\frac{n^2}{n-1}\left(x-\frac{1}{n}\right) &  x\in
\left[\frac{1}{n},1\right] \\
1 & x\in (1,\infty); \end{cases}\]
see  Figure~\ref{f2}. Furthermore, we define~$x^*:=1$ to obtain
\[h_2(x^*)=\frac{n^2-(n+1)}{-n}<0.\]
 
\begin{figure}[htbp]
 \centering
 \includegraphics[width=0.5\linewidth]{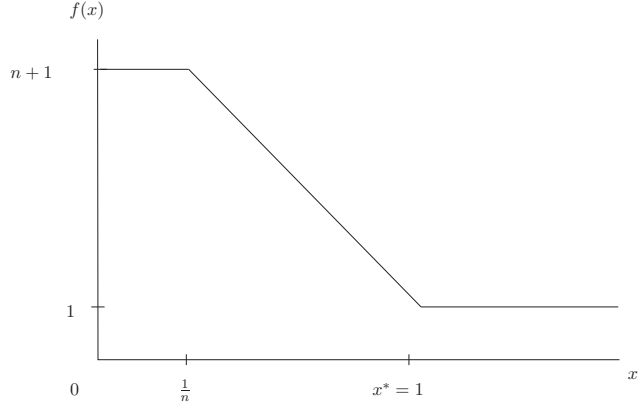}
 %\scalebox{0.7}{\includegraphics{Grafik3.eps}}  
 \caption{A shape function~$f$ for which the function~$h_2$
is not one-to-one.}
 \label{f2}
\end{figure}
 
The intuition why Theorem \ref{prop1} can be applied to this LOB shape ($f$ is decreasing), but Theorem \ref{prop3} cannot be used, is the following: For the first trade $\xi^{(2)}_0$ from (\ref{x_0 condition 2}) we might get $D_{t_{n+1}}=\e F^{-1}(\xi^{(2)}_0) \geq 1$, i.e. there are only few new shares from the resilience effect since $f(x)$ is low for $x \geq 1$. But this $\xi^{(2)}_0$ would not be optimal \footnote{Take e.g. $n=2$ and $\e=1/2$. Then for $X_0=N+\frac{9}{2}$ we get from (\ref{x_0 condition 2}) $\xi^{(2)}_0=\frac{7}{2}$, $D_{t_{n+1}}=1$ and $\xi^{(2)}_1=...=\xi^{(2)}_{N-1}=1$, $\xi^{(2)}_N=2$. The corresponding cost are higher than for the alternative strategy $\overline{\xi}^{(2)}_0=\frac{5}{2}$,  $\overline{\xi}^{(2)}_1=...=\overline{\xi}^{(2)}_{N-1}=1$, $\overline{\xi}^{(2)}_N=3$.}. We cannot have this phenomenon in Model 1 because there the resilience is proportional to the volume consumed by the large investor.
}\end{example}

\begin{remark}{\rm (Continuous-time limit of the optimal
strategy).}\label{rem_lim_model2} As in Remark~\ref{rem_lim_model1}, we can study
the asymptotic behavior of the optimal strategy as the number $N$ of trades in
$]0,T[$ tends to infinity. 
  First, we can check that  $h_2$ converges pointwise to
  $$h_2^\infty(x):=x(1+\frac{f(x)}{f(x)+xf'(x)}),$$
 and that $N(y-F(e^{-\rho \tau}
  F^{-1}(y)))$ tends to $\rho T F^{-1}(y)f(F^{-1}(y))$, provided that~$f$
  is continuously differentiable. Now, suppose that the equation
  $$F^{-1}(X_0-\rho T F^{-1}(y)f(F^{-1}(y)))=h_2^\infty(F^{-1}(y)) $$
  has a unique solution on $]0,X_0[$, which we will call~$\xi^{(2),\infty}_0$.
We can 
  check that~$\xi^{(2),\infty}_0$ is the only one possible limit for  a
subsequence of
  $\xi^{(2)}_0$, and it is therefore its limit. We can then show that
  $N\xi^{(2)}_1$ converges to $\rho
  TF^{-1}(\xi^{(2),\infty}_0)f(F^{-1}(\xi^{(2),\infty}_0))$ and $\xi^{(2)}_N$ 
  to 
$$\xi^{(2),\infty}_T:=X_0-\xi^{(2),\infty}_0-\rho T
F^{-1}(\xi^{(2),\infty}_0)f(F^{-1}(\xi^{(2),\infty}_0)) .$$
Thus, in
  the continuous-time limit, the optimal
strategy consists in an initial block order of $\xi^{(2),\infty}_0$  shares at
time~$0$, continuous buying at the constant rate $\rho 
F^{-1}(\xi^{(2),\infty}_0)f(F^{-1}(\xi^{(2),\infty}_0))$
during $]0,T[$, and a final block order of 
  $\xi^{(2),\infty}_T$ shares at time~$T$.
 
\end{remark}
%*******************************************************************************************
\section{ Closed form solution for block-shaped LOBs and additional permanent
impact.}\label{Sec4} In this first example section, we consider a block-shaped LOB
corresponding to a constant shape function~$f(x)\equiv q$ for some $q>0$. In this case,
there is no difference between Models 1 and 2.
 Apart from our more general dynamics for $A^0$, the main  difference to the market
impact model introduced by Obizhaeva and Wang~\cite{ow} is that, for the moment, we do not
consider a permanent  impact of market orders. In Corollary~\ref{Cor2}, we will see,
however, that  our results yield a closed-form solution even in the case  of nonvanishing
permanent impact.

By applying either
Theorem~\ref{prop1} or Theorem~\ref{prop3} we obtain the following Corollary. %; a detailed
%proof can be found in Section~\ref{App_OW}.
 
\medskip
 
\begin{corollary}\label{cor5}{\rm (Closed-form solution for block-shaped LOB).}{\\
In a block-shaped LOB, the unique optimal strategy $\xi^*$ is  
\begin{equation}\label{explicit strategy}
\xi^*_0=\xi^*_N=\frac{X_0}{(N-1)(1-\e)+2} \hspace{0.3 cm} \mbox{and} \hspace{0.3 cm}
\xi^*_1=\dots=\xi^*_{N-1}=\frac{X_0-2\xi^*_0}{N-1}.
\end{equation}
}
\end{corollary}

\medskip

The preceding result extends~\cite[Proposition 1]{ow} in several aspects. First,
we do not focus on the Bachelier model but admit arbitrary martingale dynamics for
our unaffected best ask price $A^{{0}}$. Second,  only static,
deterministic buy order strategies are considered in~\cite{ow}, while we here allow
our admissible strategies to be adapted and to 
include sell orders. Since,
\emph{a posteriori}, our optimal strategy turns out to be deterministic and
positive, it is clear that it must coincide with the optimal strategy
from~\cite[Proposition 1]{ow}. Our strategy~(\ref{explicit strategy})  therefore also 
provides an
\emph{explicit closed-form solution} of the recursive scheme obtained in~\cite{ow}. We
recall this recursive scheme  in~(\ref{recursive strategy}) below.

On the other hand, Obizhaewa and Wang~\cite{ow} allow for an additional \emph{permanent
impact} of market orders. Intuitively, in a block-shaped LOB with $f\equiv q>0$, the
permanent impact of a market order
$x_t$  means that only a certain part of the impact of $x_t$ decays to zero, while
the remaining part remains forever present in the LOB. More precisely, the impact of an
admissible buy order strategy $\xi$ on the extra spread $D^A$ is given by the
dynamics
\begin{equation}\label{permanentimpactdynamics}
D^A_t=\lambda\sum_{t_k<t}\xi_{k}+\sum_{t_k<t}\kappa e^{-\rho(t-t_k)}\xi_{k},
\end{equation} 
where $\lambda<1/q$ is a constant quantifying the permanent impact
and 
\begin{equation}\label{kappa}\kappa:=\frac1q-\lambda
\end{equation}
is the proportion of the temporary impact.   Note that, for
$\lambda=0$, we get back our dynamics~(\ref{Model1dynamics}) and~(\ref{Model2dynamics}), due to the fact that we consider a block-shaped LOB. It
will be convenient to introduce the process $X_t$ of the still outstanding number
of shares at time $t$ when using an admissible strategy:
\begin{equation}\label{def_Xt}X_t:=X_0-\sum_{t_k<t}\xi_k.
\end{equation}
 We can now state the result by Obizhaeva and Wang.

\medskip
 
\begin{proposition}{\rm \cite[Proposition 1]{ow}} \label{propOW}{
In a block-shaped LOB with permanent impact~$\lambda$, the optimal strategy $\xi^{OW}$ in
the class of deterministic strategies is determined by the  forward 
scheme
\begin{eqnarray}\label{recursive strategy}
\xi^{OW}_n&=&\frac{1}{2}\delta_{n+1}\left [\epsilon_{n+1}X_{t_n}-\phi_{n+1}D_{t_n}\right
],\qquad n=0,\dots,N-1,\\
\xi^{OW}_N&=&X_T,\nonumber
\end{eqnarray}
where $\delta_n$, $\epsilon_n$ and $\phi_n$ are defined by the backward scheme
\begin{eqnarray}
\nonumber  \delta_{n}&:=&\Big(\frac{1}{2q}+\alpha_{n}-\beta_{n}\kappa
e^{-\rho\tau}+\gamma_{n}\kappa^2e^{-2\rho\tau}\Big)^{-1}\\
\label{recursion2} \epsilon_{n}&:=&\lambda+2\alpha_{n}-\beta_{n}\kappa e^{-\rho\tau}\\
\nonumber \phi_{n}&:=&1-\beta_{n}e^{-\rho\tau}+2\gamma_{n}\kappa e^{-2\rho\tau}.
\end{eqnarray}
with $\alpha_n$, $\beta_n$ and~$\gamma_n$ given by
\begin{eqnarray}
\alpha_N=\frac{1}{2q}-\lambda  \hspace{0.3 cm}  \mbox{and} \hspace{0.3 cm} 
\alpha_n&=&\alpha_{n+1}-\frac{1}{4}\delta_{n+1}\epsilon_{n+1}^2,\nonumber \\
\beta_N=1   \hspace{0.3 cm} \mbox{and} \hspace{0.3   
cm}\beta_n&=&\beta_{n+1}e^{-\rho\tau}+
\frac{1}{2}\delta_{n+1}\epsilon_{n+1}\phi_{n+1},\label{recursion1}\\
\nonumber  \gamma_N=0     \hspace{0.3 cm}  \mbox{and} \hspace{0.3 cm}  
\gamma_n&=&\gamma_{n+1}e^{-2\rho\tau}-\frac{1}{4}\delta_{n+1}\phi_{n+1}^2.
\end{eqnarray}
%Moreover, given by
 
}\end{proposition}
 
\medskip

It is \emph{a priori} clear that for $\lambda=0$ the explicit optimal strategy obtained in
Corollary~\ref{cor5} must coincide with the strategy $\xi^{OW}$ obtained via the recursive
scheme~(\ref{recursive strategy}) in Proposition~\ref{propOW}. To cross-check our results
with the ones in~\cite{ow}, we will nevertheless provide an explicit and independent  proof
of the following proposition. It can be found in Section~\ref{App_OW}.
 
\medskip

\begin{proposition} For $\lambda=0$, the optimal strategy~\eqref{explicit strategy} of
Corollary~\ref{cor5} solves the recursive
scheme~\eqref{recursive strategy} in Proposition~\ref{propOW}. 
\end{proposition}

\medskip

Let us now  extend our results so as to obtain the explicit solution of~(\ref{recursive
strategy}) even with  nonvanishing permanent impact. To this end, we note that the optimal
strategy
$\xi^{OW}=(\xi^{OW}_0,\dots,\xi^{OW}_N)$ is obtained in~\cite{ow} as 
the unique minimizer of the cost functional 
$$C^{\textsc{OW}}_{\lambda,q}:\R^{N+1}\to\R$$
defined by
\begin{eqnarray}
 \lefteqn{ C^{\textsc{OW}}_{\lambda,q} (x_0,\dots,x_N)}\nonumber\\
&=&A_0\sum_{i=0}^N
x_i +
\frac{\lambda}{2}\Big( \sum_{i=0}^N x_i\Big)^2 +
\kappa\sum_{k=0}^N\Big(\sum_{i=0}^{k-1} x_i e^{- \rho (k-i) \tau}
\Big)x_k  \nonumber +\frac{\kappa}{2}  \sum_{i=0}^N x_i^2,
\end{eqnarray}
where $\kappa$ is as in~\eqref{kappa}. Now
we just have to observe that
$$C^{\textsc{OW}}_{\lambda,q}(x_0,\dots,x_N)=\frac{\lambda}{2}\Big( \sum_{i=0}^N
  x_i\Big)^2 +C^{\textsc{OW}}_{0,{\kappa^{-1}}}(x_0,\dots,x_N).$$
Therefore, under the constraint $\sum_{i=0}^N  x_i = X_0$, it is equivalent to
minimize either $C^{\textsc{OW}}_{\lambda,q}$ or $C^{\textsc{OW}}_{0,{\kappa^{-1}}}$.
We already know that the optimal strategy $\xi^*$ of
Corollary \ref{cor5} minimizes $C^{\textsc{OW}}_{0,{q}}$. But $\xi^*$ is in fact
independent of
$q$. Hence, $\xi^*$ also minimizes $C^{\textsc{OW}}_{0,{\kappa^{-1}}}$ and in turn
$C^{\textsc{OW}}_{\lambda,q}$. We have therefore proved:
 
\medskip
 
\begin{corollary}\label{Cor2}The optimal strategy $\xi^*$ of
Corollary~\ref{cor5} is the unique optimal
strategy in any block-shaped LOB with permanent impact $\lambda<1/q$. In particular, it
solves the recursive scheme~\eqref{recursive strategy}.
\end{corollary}
 
\medskip
 
The last part of the assertion of Corollary~\ref{Cor2} is remarkable insofar as the
recursive scheme~\eqref{recursive strategy} depends on both $q$ and $\lambda$ whereas the optimal
strategy $\xi^*$ does not.

%*******************************************************************************************
\section{Examples.}\label{Sec5}
 
\begin{figure}[htbp]
 \centering
 \psfrag{x}{$x$}
 \psfrag{f}{$f(x)$}
 \includegraphics[width=0.5\linewidth]{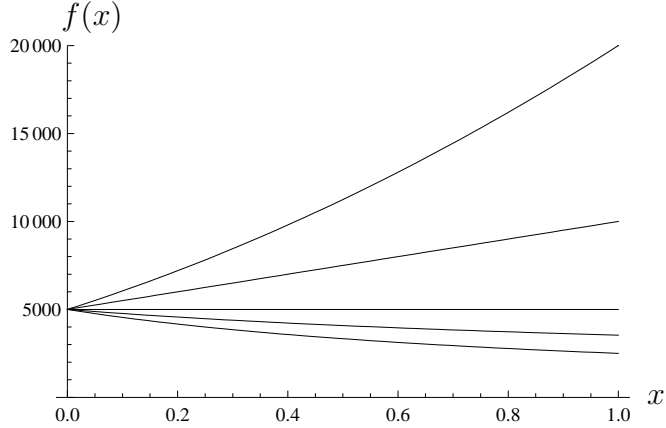}
%\scalebox{0.7}{\includegraphics{Grafik3.eps}}  
 \caption{Plots of the power law shape functions for~$q=5,000$ shares and exponent $\alpha=-2,-1,0,\frac{1}{2}$ and $1$ top down. Please note that these examples do not necessarily correspond to real-world shape functions.}
 \label{LOBplot}
\end{figure}
 
In this section, we  consider the power law family $f:\R \rightarrow \R_{>0}$ with
\begin{equation}\label{power law}
            f(x)=\frac{q}{(|x|+1)^\alpha}
\end{equation}
as example shape functions. The antiderivative of the shape function and its inverse are
$$ F(x)=\begin{cases} 
 q \log(x+1) &\text{if } \alpha=1 \\ 
 qx &\text{if } \alpha=0\\
 \frac{q}{1-\alpha} \left[(x+1)^{1-\alpha}-1 \right]& \text{otherwise }
\end{cases} 
\hspace{.5 cm} 
F^{-1}(y)=\begin{cases} 
 e^{\frac{y}{q}}-1  &\text{if } \alpha=1 \\ 
 \frac{y}{q} &\text{if } \alpha=0\\
 \left[1+(1-\alpha)\frac{y}{q} \right]^{\frac{1}{1-\alpha}}-1 & \text{otherwise }
\end{cases}$$
for positive values of $x$ and $y$. Set $F(x)=-F(-|x|)$ and $F^{-1}(y)=-F^{-1}(-|y|)$ for $x,y<0$.
 
\begin{figure}
\begin{minipage}[c]{0.5\textwidth}
\centering 
\psfrag{alpha}{$\alpha$}
\psfrag{Initial}{Initial and last trade}
\includegraphics[width=3in]{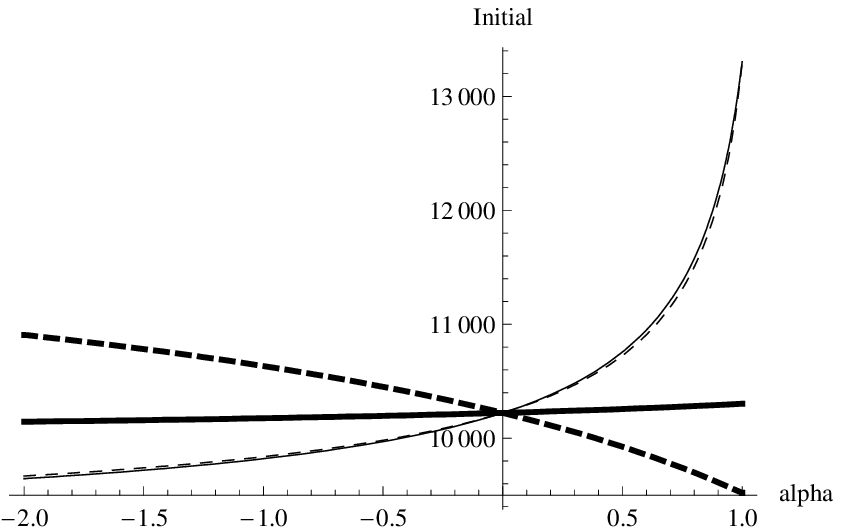}
\end{minipage}% Prozentzeichen verhindert Leerzeichen!
\begin{minipage}[c]{0.5\textwidth}
\centering 
\psfrag{alpha}{$\alpha$}
\psfrag{Intermediate}{Intermediate trades}
\includegraphics[width=3in]{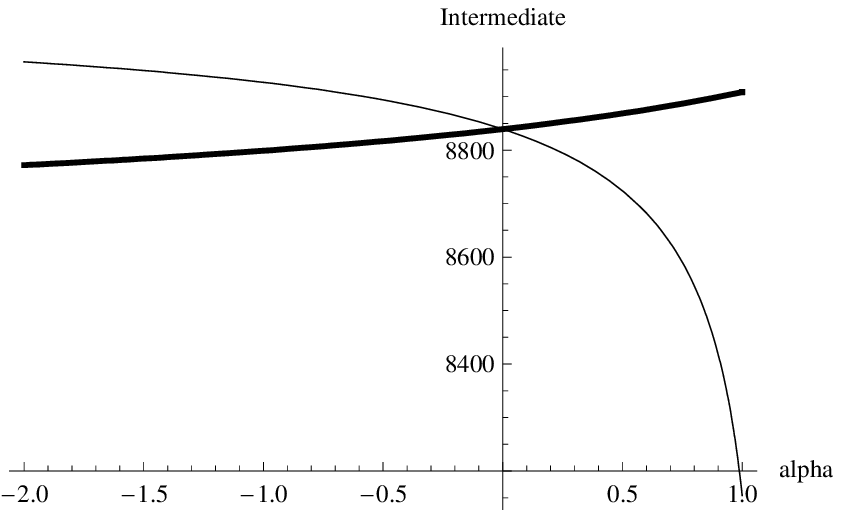}
\end{minipage}
\caption{The plots show the optimal strategies for varying exponents $\alpha$. We set $X_0=100,000$ and $q=5,000$ shares, $\rho=20$, $T=1$ and $N=10$. In the left figure we see $\xi^{(1)}_0$ (dashed and thick), $\xi^{(1)}_N$ (thick line) and $\xi^{(2)}_0$, $\xi^{(2)}_N$. The figure on the right hand side shows $\xi^{(1)}_1$ (thick line) and $\xi^{(2)}_1$.}
 \label{optimal strategies}
\end{figure}

One can easily check that the assumptions of both  Theorem~\ref{prop1} and Theorem~\ref{prop3} are satisfied for $\alpha \leq 1$. It is
remarkable that the optimal strategies (Figure \ref{optimal strategies}) vary only slightly when changing $\alpha$ or the resilience mode. This observation provides further evidence for the robustness and stability of the optimal strategy, and this time not only on a qualitative
but also on a quantitative level.

From Figure \ref{optimal strategies} one recognizes some monotonicity properties of the optimal strategies. We want to give some intutition to understand these.
Let us start with Model 1. There the dynamics of $E_t$ do not depend on the LOB shape, but solely on the strategy. Only the cost depends on $f$. We know from the constant LOB case that the optimum strategy is not sensible to the value of $f(\xi^{(1)}_0)$. This explains why there are few quantitative differences for Model 1 along the different LOB shapes. Moreover, $\xi^{(1)}_1=(1-a)\xi^{(1)}_0$ with $a:=\e$ is proportional to~$\xi^{(1)}_0$ since it is the number of shares that reappear between two trades. Therefore the optimal strategy is just a trade-off between $\xi^{(1)}_0$ and $\xi^{(1)}_N$. When $f$ is increasing (decreasing), the first trade is relatively more (less) expansive compared to the last one. This explains that $\xi^{(1)}_0 < \xi^{(1)}_N$ for $\alpha<0$ and $\xi^{(1)}_0 > \xi^{(1)}_N$ for $\alpha >0$. With `relatively' we mean 'with respect to the constant LOB case' ($\alpha=0$) where $\xi_0 = \xi_N$.

For Model 2 the dynamics of $E_t$ do depend on the shape function, which explains
more substantial variations according to $f$. Here the main idea is to realize that,
for increasing (decreasing) shape functions, resilience of the volume is stronger (weaker)  in comparison to Model 1. {Indeed, we have then
  $x-F(aF^{-1}(x))\ge x(1-a) $ (resp. $x-F(aF^{-1}(x))\le x(1-a) $).} Therefore $\xi^{(1)}_1 < \xi^{(2)}_1$ ($\xi^{(1)}_1 > \xi^{(2)}_1$) and the discrete trades $\xi^{(2)}_0$ and $\xi^{(2)}_N $ are lower (higher) as in Model 1. These effects are the more pronounced the steeper the LOB shape. Furthermore, there is the tendency that $\xi^{(2)}_0 \approx \xi^{(2)}_N$. On the one hand, the same argument as in Model 1 suggests $\xi^{(2)}_0 < \xi^{(2)}_N$ for increasing $f$. But on the other hand, for an increasing shape function the number of reappearing shares grows disproportionately in the initial trade which favors the initial trade being higher than the last trade. These two effects seem to counterbalance each other.

\begin{remark} Taking the special LOB shape  $f(x)= \frac{q}{\sqrt{1+ \mu |x|}}$,
  $q>0$ and $\mu \ge 0$ we can solve explicitly the optimal strategy in Model~1
  from Theorem~\ref{prop1}. The optimal initial trade is given by 
\begin{eqnarray*}\xi^{(1)}_0&=& \frac{1+a+N(1-a)(1+(\mu/2q) X_0)}{(\mu/2q)
  (N^2(1-a)^2-(1+a+a^2)) }\\& &-\frac{ \sqrt{(N+1-a(N-1))^2 +(\mu/q) X_0[N (1-a^2) +(1+a+a^2) (1+ (\mu/4q) X_0)]}}{(\mu/2q)
  (N^2(1-a)^2-(1+a+a^2)) },
\end{eqnarray*}
and we can show that it is increasing with respect to
the parameter $\mu$ that tunes the slope of the LOB.
\end{remark}

\begin{appendix}
\section{Reduction to the case of
deterministic strategies.} \label{ReductionSection}
 
In this section,  we prepare for the proofs of
Theorems~\ref{prop1} and~\ref{prop3} by reducing the  minimization of the cost functional
$$\cC(\xi)=\mathbb{E}\Big[\,\sum_{n=0}^N\pi_{t_n}(\xi_n)\,\Big]
$$ 
 with
respect to all admissible strategies $\xi$ to 
the minimization of certain cost functions $C^{(i)}:\R^{N+1}\to\R$, where $i=1,2$ refers to
the model under consideration.

To this end, we introduce simplified versions of the model dynamics by collapsing
the bid-ask spread into a single value. More precisely, for any admissible
strategy $\xi$, we introduce a new pair of processes $D$ and $E$ that react on both
sell and buy orders according to the following dynamics.
\begin{itemize}
\item We have $E_0=D_0=0$ and 
\begin{equation}\label{E->D2}
E_t=F(D_t)\qquad\text{and}\qquad D_t=F^{-1}(E_t).
\end{equation}
\item For $n=0,\dots,N$, regardless of the sign of $\xi_n$,
\begin{equation}\label{EDjumps} E_{t_n+}=E_{t_n}+\xi_n  
\qquad\text{and}\qquad
D_{t_n+}=F^{-1}\left(\xi_n+F\left(D_{t_n}\right)\right).
\end{equation}
\item For $k=0,\dots,N-1$,
\begin{equation}\label{DEdynamics}\begin{split} E_{t_{k+1}}&=\e E_{t_k+}
\qquad\text{in Model 1,}\\  D_{t_{k+1}}&=\e D_{t_k+} \qquad\text{in Model 2.}
\end{split}
\end{equation}
\end{itemize}
The values of $E_t$ and $D_t$ for $t\notin\{t_0,\dots,t_N\}$ will not be needed in
the sequel.
Note that $E=E^A$ and $D=D^A$ if $\xi$ consists only of buy orders, while $E=E^B$
and $D=D^B$ if $\xi$ consists only of sell orders. In general, we will only have
\begin{equation}\label{EBEEA}
E^B_t\le E_t\le E^A_t\qquad \text{and}\qquad D^B_t\le D_t\le D^A_t.
\end{equation}

We now introduce the \emph{simplified price} of $\xi_n$ at time $t_n$ by
\begin{equation}\label{}\bbar\pi_{t_n}(\xi_n):=
A^{{0}}_{t_n}\xi_n+\int_{D_{t_n}}^{D_{t_n+}}xf(x)\,dx,
\end{equation}
regardless of the sign of $\xi_n$. Using \eqref{EBEEA} and \eqref{B0t}, we easily
get 
\begin{equation}\label{buyorderprice}
\bbar\pi_{t_n}(\xi_n)\le\pi_{t_n}(\xi_n)\qquad\text{with
equality if $\xi_k\ge0$ for all $k\le n $.}
\end{equation}
The \emph{simplified price functional} is defined as
$$\bbar\cC(\xi):=\mathbb{E}\Big[\,\sum_{n=0}^N\bbar\pi_{t_n}(\xi_n)\,\Big].
$$ 
We will show that, in Model $i\in\{1,2\}$, the simplified price functional
$\bbar\cC$  has a unique minimizer, which coincides with the corresponding optimal
strategy $\xi^{(i)}$ as described in the respective theorem. We will also show
that $\xi^{(i)}$ consists only of buy orders, so that \eqref{buyorderprice} will
yield $\cC(\xi^{(i)})=\bbar \cC(\xi^{(i)})$.  Consequently, $\xi^{(i)}$ must be
the unique minimizer of $\cC$.

Let us now reduce the minimization of $\bbar\cC$ to the minimization of
functionals $C^{(i)}$ defined on deterministic strategies. To this end, let us use
the notation 
\begin{equation}\label{}
\text{$X_t:=X_0-\sum_{t_k<t}\xi_k$ for $t\le T$ and
$X_{t_{N+1}}:=0$.}
\end{equation}
The accumulated simplified price of an
admissible strategy $\xi$ is
$$\sum_{n=0}^N\bbar\pi_{t_n}(\xi_n)=
\sum_{n=0}^NA^{{0}}_{t_n}\xi_n+\sum_{n=0}^N\int_{D_{t_n}}^{D_{t_n+}}xf(x)\,dx.
$$ 
 Integrating by parts yields
\begin{equation}\label{systemeq}\sum_{n=0}^NA^{{0}}_{t_n}\xi_n=
-\sum_{n=0}^{N}A^{{0}}_{t_n}(X_{t_{n+1}}-X_{t_n})=X_0A_0+
\sum_{n=1}^{N}X_{t_n}(A^{{0}}_{t_{n}}-A^{{0}}_{t_{n-1}}).
\end{equation}
Since $\xi$ is admissible, $X_t$ is a bounded predictable process. Hence, due to the
martingale property of the unaffected best ask process $A^{{0}}$, the expectation
of \eqref{systemeq} is equal to $X_0A_0$. 
 
Next, observe that, in each Model $i=1,2$, the simplified extra spread process $D$
evolves deterministically once the values
$\xi_0,\xi_1(\omega),\dots,\xi_N(\omega)$ are given. Hence, there exists a
deterministic function
$C^{(i)}:\R^{N+1}\to\R$ such that
\begin{equation}\label{Ci}
\sum_{n=0}^N\int_{D_{t_n}}^{D_{t_n+}}xf(x)\,dx=C^{(i)}(\xi_0,\dots,\xi_N).
\end{equation}
It follows that 
$$\bbar\cC(\xi)=A_0X_0+\mathbb{E}\big[\,C^{(i)}(\xi_0,\dots,\xi_N)\,\big].
$$
We will show in the respective Sections~\ref{Model1proofsection}
and~\ref{Model2proofsection} that the functions $C^{(i)}$, $i=1,2$, have unique
minima within 
the set 
\[\Xi:=\Big\{\left(x_0,\dots,x_N\right) \in \R^{N+1}\,\big|\,\sum_{n=0}^Nx_n=X_0\Big\},\]
and that these minima coincide with the values of the optimal strategies $\xi^{(i)}$ as
provided in  Theorems~\ref{prop1}
and~\ref{prop3}. This concludes the reduction to the case of deterministic strategies. 
We will now turn to the minimization of the functions $C^{(i)}$ over $\Xi$. To
simplify the exposition, let us introduce the 
following shorthand notation in the sequel:
\begin{equation}\label{Delta}{{a}}:=\e.
\end{equation}

\section{The optimal strategy in Model 1.}\label{Model1proofsection}
 
In this section, we will minimize the function $C^{(1)}$ of \eqref{Ci} over the
set $\Xi$ of all deterministic strategies and thereby complete the proof of
Theorem~\ref{prop1}. To this end, recall first the definition of the two processes 
$E$ and
$D$ as given in
\eqref{E->D2}--\eqref{DEdynamics}. Based on their Model 1 dynamics, we will now
obtain  a formula of the cost function
$C^{(1)}$ of
\eqref{Ci} in terms of the functions $F$ and $\widetilde{F}$. It will be convenient to
introduce also the function
\begin{equation}\label{Gfunction}G(y):=\widetilde{F}\left(F^{-1}(y)\right).
\end{equation}
Then we have for any deterministic strategy $\xi=(x_0,\dots,x_N)\in\Xi$ that
\begin{eqnarray}
C^{(1)}(x_0,\dots,x_N)&=&\sum_{n=0}^N \int_{D_{t_n}}^{D_{t_n+}}x f(x)dx \nonumber \\
&=&\sum_{n=0}^N  
\left(\widetilde{F}\left(F^{-1}\left(E_{t_n+}\right)\right) -
\widetilde{F}\left(F^{-1}\left(E_{t_n}\right)\right)\right) \nonumber \\
       &=&\sum_{n=0}^N 
\big( G\left(E_{t_n}+x_n\right) - G\left(E_{t_n}\right)
\big) \label{GEeq} \\
&=& \phantom{+} G\left(x_0\right)-G\left(0\right)\nonumber\\
\nonumber                 &&+G\left({{a}}x_0 +x_1\right)-G\left({{a}}x_0\right)\\
\nonumber                 &&+G\left({{a}}^2x_0 + {{a}}x_1+x_2\right)-G\left({{a}}^2x_0
+{{a}}x_1 \right)\\
\label{C_0}               &&+\dots\\
\nonumber                 &&+G\left({{a}}^Nx_0 +\dots+x_N\right)-G\left({{a}}^Nx_0
+\dots+  {{a}}x_{N-1} 
\right).
\end{eqnarray}
The derivative of~$G$ is
\begin{equation}\label{G derivative}
G'(y)=\widetilde{F}'\left(F^{-1}(y)\right)(F^{-1})'(y)=F^{-1}(y) f\left(F^{-1}(y)\right)   
\frac{1}{f(F^{-1}(y))}=F^{-1}(y).
\end{equation}
Hence, $G$ is twice continuously differentiable, positive and convex. The cost
function  $C^{(1)}$ is also twice continuously differentiable.

\medskip

\begin{lemma}\label{lemma2}{We have $C^{(1)}(x_0,\dots,x_N){\longrightarrow}+ \infty$ for
$|\xi|\to\infty$, and therefore there exists a local minimum of~$C^{(1)}$ in~$\Xi$.}
  \end{lemma}
 
\Proof Using~\eqref{G derivative} and the fact that $F^{-1}(yx)$ is increasing, we get
that  for all~$y \in \R$ and~$c \in (0,1]$
\begin{equation}\label{G inequality}
G(y)-G(c y)\geq (1-c) \cdot|F^{-1}(c y)|\cdot|y|.
\end{equation}
 Let us rearrange the sum in~(\ref{C_0}) in order to use inequality~(\ref{G inequality}). We obtain
\begin{eqnarray*}
\lefteqn  { C^{(1)}(x_0,\dots,x_N)}\\
\nonumber  &=& G\big({{a}}^Nx_0 + {{a}}^{N-1}x_1+\dots+x_N
\big)-G\left(0\right)\\
  \nonumber  &&+ \sum_{n=0}^{N-1} \Big[G\big({{a}}^{n} x_0+\dots+x_n\big)
-G\big({{a}}({{a}}^{n} x_0+\dots+x_n)\big)\Big] \\
\nonumber  & \geq & G\big({{a}}^Nx_0 +{{a}}^{N-1}x_1 +\dots+x_N
\big)-G\left(0\right)\\
\nonumber  &&+ (1-{{a}})\sum_{n=0}^{N-1}\left|F^{-1}\big({{a}}({{a}}^{n}
x_0+\dots+x_n)\big)\right|
\left|{{a}}^{n} x_0+\dots+x_n \right|.
\end{eqnarray*}
Let us denote by $T_1:\R^{N+1} \rightarrow \R^{N+1}$ the linear mapping
$$T_1(x_0,\dots,x_n)=\big(x_0,{{a}} x_0+x_1,\dots,{{a}}^{N} x_0+x_1
{{a}}^{N-1}+\dots+x_N
\big).$$ 
It is non trivial and therefore the norm of $T_1(x_0,\dots,x_N)$ tends to infinity as 
the norm of its argument goes to infinity. Because $F$ is
unbounded, we know that both~$G(y)$ and
$|F^{-1}({{a}} y)||y|$
    tend to infinity for $|y| \rightarrow \infty$. Let us introduce
    $$H(y)=\min(G(y),|F^{-1}({{a}} y)||y|).
$$
Then also $H(y){\longrightarrow} + \infty$ for $|y|\to\infty$, and we conclude that
    $$C^{(1)}(x_0,\dots,x_N) \ge (1-{{a}}) H(|T_1(x_0,\dots,x_N)|_\infty)
-G(0)
     , $$
where $|\cdot|_\infty$ denotes the $\ell^\infty$-norm on $\R^{N+1}$. Hence, the assertion
follows.
\hfill  \sq

\medskip
 
We now consider Equation \eqref{x_0 condition 1} in Theorem~\ref{prop1}, which we recall
here for the convenience of the reader:
$$ F^{-1}\left(X_0-N x_0\left(1-{{a}}\right)\right)=\frac{h_1(x_0)}{1-a}.
$$
 This equation is solved by $x_0$ if and only if $x_0$ is a zero of the function
\begin{equation}\label{hath1}\hat{h}_1(y):=h_1(y)- (1-{{a}} )F^{-1}\big(X_0-N y
 (1-{{a}} )\big).
\end{equation}
 
\medskip
 
\begin{lemma}\label{uniqueness} Under the assumptions of the Theorem~\ref{prop1},
  $\hat h_1$ has at most one
  zero~$x_0$, which, if it exists,  is necessarily positive.
\end{lemma}
 
\Proof  It is sufficient to show that~$\hat{h}_1$ is strictly increasing.
We know that~$h_1(0)=0$, $h_1(y)>0$ for~$y>0$, and~$h_1$ is continuous and one-to-one. 
Consequently, $h_1$ must be strictly increasing and therefore
\[\hat{h}_1'(y)=h_1'(y)+\frac{N  ({{a}}-1 )^2}{f\big(F^{-1}\left(X_0+N y
\left({{a}}-1\right)\right)\big)}>0. \] 
Furthermore, if there exists a solution $x_0$, then it must be positive since
\[ \hat{h}_1(0)= ({{a}}-1 )  F^{-1}(X_0)<0. \]
\hfill  \sq

%% \Proof  It is sufficient to show that~$\hat{h}_1$ is strictly increasing.
%% We know that~$h_1(0)=0$, $h_1(y)>0$ for~$y>0$, and~$h_1$ is continuous and one-to-one. 
%% Consequently, $h_1$ must be strictly increasing and therefore
%% \[\hat{h}_1'(y)=h_1'(y)+\frac{N  ({{a}}-1 )^2}{f\big(F^{-1}\left(X_0+N y
%% \left({{a}}-1\right)\right)\big)}>0. \] 
%% Furthermore, if there exists a solution $x_0$, then it must be positive since
%% \[ \hat{h}_1(0)= ({{a}}-1 )  F^{-1}(X_0)<0. \]
%% \hfill  \sq
%% \medskip
Theorem~\ref{prop1} will now follow by combining the following proposition with the
arguments explained in Section \ref{ReductionSection}.

\begin{proposition} The function $C^{(1)}:\Xi\to\R$ has the strategy $\xi^{(1)}$
from Theorem~\ref{prop1} as its unique minimizer.
Moreover, the components of $\xi^{(1)}$ are all strictly positive.
\end{proposition}

\Proof
Thanks to Lemma~\ref{lemma2},  there is at
least one optimal strategy $\xi^*=(x_0^*,\dots,x_N^*)\in\Xi$, and standard results give the
existence of a Lagrange multiplier~$\nu \in \R$ such that 
\[\frac{\partial}{\partial
x_i}C^{(1)}(x_0^*,\dots,x_N^*)=\nu\qquad\text{for~$i=0,\dots,N$.} \]
 Now we use the form of~$C^{(1)}$ as given in~(\ref{C_0}) to obtain the following 
relation between the partial derivatives of~$C^{(1)}$ for~$i=0,\dots,N-1$:
\begin{eqnarray*}
\frac{\partial}{\partial x_i}C^{(1)}(x_0,\dots,x_N)&=&{{a}} \left[\frac{\partial}{\partial x_{i+1}}C^{(1)}(x_0,\dots,x_N)-
G'\left({{a}}({{a}}^ix_0+\dots+x_i)\right)  \right]\\
&+&G'\left({{a}}^ix_0+\dots+x_i\right)
\end{eqnarray*}
Recalling~(\ref{G derivative}), we obtain
\[ h_1\left({{a}}^ix_0^*+\dots+x_i^*\right)=\nu \left(1-{{a}}\right) \hspace{0.5 cm}
\mbox{for $i=0,\dots,N-1$}. \] 
Since $h_1$ is one-to-one we must have
\begin{eqnarray}\label{i}
\nonumber   x^*_0&=&h_1^{-1}\left(\nu\left(1-{{a}}\right)\right)\\
x^*_i&=&x^*_0\left(1-{{a}}\right)  \hspace{0.5 cm} \mbox{for $i=1,\dots,N-1$}\\
\nonumber x^*_N&=&X_0-x^*_0-(N-1)x^*_0\left(1-{{a}}\right).
\end{eqnarray}
Note that these equations link all the trades to the initial trade~$x_0$. Due to the
dynamics~\eqref{EDjumps} and \eqref{DEdynamics}, it follows  that the process $E$
of
$\xi^*$ is given by 
\begin{equation}\label{EoptstratModel1}
E_{t_n}={{a}} \left({{a}}x_0
+x_0\left(1-{{a}}\right)\right)={{a}}x_0.
\end{equation}
Consequently, by
\eqref{GEeq},
\begin{eqnarray*}C^{(1)}(x^*_0,\dots,x^*_N) &=& G(x^*_0)-G(0)+(N-1)\big[G\left({{a}}x^*_0
+x^*_0(1-{{a}})\right)-G({{a}}x^*_0 )\big]\\
\nonumber  && +G\big({{a}}x^*_0 +X_0-x^*_0-(N-1)x^*_0(1-{{a}})\big)-G(x^*_0 {{a}})\\
&=&N\big[G(x^*_0)-G(x^*_0 {{a}})\big]+G\big(X_0+N x^*_0({{a}}-1)\big)-G(0)\\
&=:& C^{(1)}_0(x^*_0).
\end{eqnarray*}
It thus remains to minimize the function $C^{(1)}_0(y)$ with respect to $y$. Thanks to
the existence of an optimal strategy in~$\Xi$ for $C^{(1)}$, we know that $C^{(1)}_0(y)$
has at least one local minimum.  Differentiating  with respect to~$y$ gives
\begin{eqnarray}
\nonumber     \frac{\partial C^{(1)}_0(y)}{\partial y}&=&
N \left[F^{-1}(y)-{{a}} F^{-1}\left( {{a}}y\right)
                     +\left({{a}}-1\right) F^{-1}\left(X_0+N y\left({{a}}-1\right)\right)\right
]\\
&=&N\hat h_1(y).\label{C derivative 1}
                 \end{eqnarray}
 Lemma~\ref{uniqueness} now implies that $C^{(1)}_0$ can only have one local minimum,
which is also positive if it exists. This local minimum must hence be equal to $x_0^*$,
which establishes both the uniqueness of the optimal strategy as well as our
representation.

Finally, it remains to prove that all market orders in the optimal strategy  are
strictly positive. Lemma~\ref{uniqueness} gives $\xi_0^{(1)}=x^*_0>0$ and
then~\eqref{i} gives
$\xi_n^{(1)}=x^*_n>0$ for
$n=1,\dots,N-1$. As for the final market order, using the facts that \eqref{C derivative 1}
vanishes at $y=x_0^*$ and $F^{-1}$ is strictly increasing gives
\begin{eqnarray*}0&=&F^{-1}(x^*_0)-aF^{-1}(ax_0^*)-(1-a)F^{-1}(ax_0^*+x^*_N)\\
&>&(1-a)\big[F^{-1}(ax_0^*)-F^{-1}(ax_0^*+x^*_N)\big],
\end{eqnarray*}
which in turn implies $x_N^*>0$.
\hfill
\sq

\medskip

%**********************************************************************************************************
\section{The optimal strategy in Model 2.}\label{Model2proofsection}
In this section, we will minimize the function $C^{(2)}$ of \eqref{Ci} over the
set $\Xi$ of all deterministic strategies and thereby complete the proof of Theorem~\ref{prop3}. To this end, recall first that the definitions of $D$ and $E$  are given by
\eqref{E->D2}--\eqref{DEdynamics}. Based on their Model 2 dynamics, we
will now obtain  a formula of the cost function
$C^{(2)}$ of
\eqref{Ci} in terms of the functions $F$, $\widetilde{F}$, and $G$, where $G$ is as in
\eqref{Gfunction}. For any deterministic strategy $\xi=(x_0,\dots,x_N)\in\Xi$,
\begin{eqnarray}
C^{(2)}(x_0,\dots,x_N)&=&\sum_{n=0}^N \int_{D_{t_n}}^{D_{t_n+}}x f(x)dx \nonumber \\
&=& \sum_{n=0}^N  \left(G\left(x_n+F\left(D_{t_n}\right)\right) -
\widetilde{F}\left(D_{t_n}\right)\right).\label{costG}
\end{eqnarray}
 
\medskip
{We now state three technical lemmas that will allow to get the optimal strategy. }
\begin{lemma}\label{lemma4}{We have $C^{(2)}(x_0,\dots,x_N){\longrightarrow}+ \infty$
for $|\xi|\to\infty$, and therefore there exists a local minimum of~$C^{(2)}$ in~$\Xi$.}
\end{lemma}
 
\Proof We rearrange the sum in~(\ref{costG}):
\begin{eqnarray}\nonumber
 & C^{(2)}(x_0,\dots,x_N)&= \widetilde{F}\left({{a}} F^{-1}(x_N+F(D_{t_N}))\right)\\
  \nonumber     && + \sum_{n=0}^{N}\left[ \widetilde{F}\left(F^{-1}(x_n+F(D_{t_n}))\right)  
    -\widetilde{F}\left({{a}} F^{-1}(x_n+F(D_{t_n}))\right)\right] \\
  \label{C_0 spread}&& \ge  \sum_{n=0}^{N}\left[
\widetilde{F}\left(F^{-1}(x_n+F(D_{t_n}))\right)  
    -\widetilde{F}\left({{a}} F^{-1}(x_n+F(D_{t_n}))\right)\right]  .
\end{eqnarray}
For the terms in~(\ref{C_0 spread}), we have the lower bound
\begin{align*}
\widetilde{F}(z)-\widetilde{F}({{a}} z)=\left | \int_{{{a}} z}^z x f(x) dx \right |
    \geq \frac{1}{2} (1-{{a}}^2)z^2 \inf_{\widetilde{z} \in [{{a}} z,z]}f(\widetilde{z}) \ge 0.
  \end{align*}
Let 
$$H(y)=\frac{1}{2} (1-{{a}}^2)F^{-1}(y)^2 \inf_{x \in [{{a}} F^{-1}(y),F^{-1}(y)]}f(x).$$
Then 
we have $H(y){\longrightarrow} + \infty$ for $|y|\to\infty$, due  to~(\ref{explosion
assumption}) and~\eqref{F unbounded}. Besides, we have $$ C^{(2)}(x_0,\dots,x_N) \ge
H(|T_2(\xi)|_\infty)$$ 
where $|\cdot|_\infty$ denotes again the $\ell^\infty$-norm on $\R^{N+1}$, and $T_2$ is
the (nonlinear) transformation
$$T_2(\xi)=\big(x_0,x_1+ F^{-1}(D_{t_1}),\dots,x_N+F^{-1}(D_{t_N}) \big).$$
It is sufficient to show that $|T_2(\xi)|_\infty  {\longrightarrow}  \infty$ when
$|\xi|\to\infty$. To prove this, we suppose by way of contradiction that there is a
sequence
$\xi^k$ such that
$|\xi^k |_
\infty
\longrightarrow 
\infty$ and $T_2(\xi^k)$ stays bounded. Then, all coordinates in the sequence
$(T_2(\xi^k))_k$ are bounded, and in particular $(x_0^k)_k$ is a bounded sequence.
Therefore,
$D_{t_1}^k={{a}} F^{-1}( x_0^k)$ is also a bounded sequence.  The second coordinate
$x_1^k+ F^{-1}(D^k_{t_1})$ being also bounded, we get that $(x_1^k)_k$ is a bounded
sequence. In that manner, we get that $(x_n^k)_k$ is a bounded sequence for
any
$n=0,\dots,N$, which is the desired contradiction.
\hfill\sq

\begin{lemma}\label{partial_der}{\rm (Partial derivatives of~$C^{(2)}$).}\\
{
We have the following recursive scheme for the derivatives of~$C^{(2)}(x_0,\dots,x_N)$ \\for $i=0,\dots, N-1$:
\begin{eqnarray}\label{lembeh}
\qquad\frac{\partial}{\partial
x_i}C^{(2)}=F^{-1}\left(x_i+F(D_{t_i})\right)+\frac{{{a}}
f\left(D_{t_{i+1}}\right)}{f\left(F^{-1}\left (x_i+F(D_{t_i})\right)\right)} \left
[\frac{\partial}{\partial x_{i+1}}C^{(2)}-D_{t_{i+1}}\right].
\end{eqnarray}
}
\end{lemma}
 
\Proof 
From~(\ref{delta dynamic}) we get the following scheme for $D_{t_n}$ for a
fixed~$n\in \{1,\dots,N\}$:
$$\begin{array}{c c c c}
 D_{t_n}&&&\\
 \| &&&\\
 {{a}} F^{-1}(x_{n-1}+ & F(D_{t_{n-1}})) &&\\
 & \| &&\\
 & \ddots &&\\
 & {{a}} F^{-1}(x_{i+1}+ & F(D_{t_{i+1}})) &\\
 && \| &\\
 && {{a}} F^{-1}(x_i+ & F(D_{t_i})) \\
 &&& \| \\
 &&& \ddots \\
 &&& {{a}} F^{-1}( x_0).
 \end{array}$$
Therefore the following relation holds for the partial derivatives of $D_{t_n}$:
\begin{equation}\label{deltatrick}
\frac{\partial}{\partial x_i}D_{t_n}=
\frac{{{a}} f(D_{t_{i+1}})}{f\left(F^{-1}\left(x_i+F(D_{t_i})\right)\right)}\frac{\partial}{\partial
x_{i+1}}D_{t_n},\qquad i=0,\dots,n-2.
\end{equation}
Furthermore, according to~(\ref{costG}) and~(\ref{G
derivative}),
\begin{eqnarray}\label{dx_i C_0}
\lefteqn{\frac{\partial}{\partial
x_i}C^{(2)}=F^{-1}\left(x_i+F(D_{t_i})\right)+}\\
\nonumber&&\qquad \qquad +
\sum_{n=i+1}^N f(D_{t_n})\frac{\partial}{\partial x_i}D_{t_n} 
              \left[ F^{-1}\left(x_n+F(D_{t_n})\right)-D_{t_n} \right]
\end{eqnarray}
for~$i=0,\dots,N$.  Combining~(\ref{dx_i C_0}) and~(\ref{deltatrick}) yields~(\ref{lembeh}). Note that~(\ref{deltatrick}) is only valid up to $i=n-2$.
\hfill\sq

%% \Proof 
%% From~(\ref{delta dynamic}) we get the following scheme for $D_{t_n}$ for a
%% fixed~$n\in \{1,\dots,N\}$:
%% $$\begin{array}{c c c c}
%%  D_{t_n}&&&\\
%%  \| &&&\\
%%  {{a}} F^{-1}(x_{n-1}+ & F(D_{t_{n-1}})) &&\\
%%  & \| &&\\
%%  & \ddots &&\\
%%  & {{a}} F^{-1}(x_{i+1}+ & F(D_{t_{i+1}})) &\\
%%  && \| &\\
%%  && {{a}} F^{-1}(x_i+ & F(D_{t_i})) \\
%%  &&& \| \\
%%  &&& \ddots \\
%%  &&& {{a}} F^{-1}( x_0).
%%  \end{array}$$
%% Therefore the following relation holds for the partial derivatives of $D_{t_n}$:
%% \begin{equation}\label{deltatrick}
%% \frac{\partial}{\partial x_i}D_{t_n}=
%% \frac{{{a}} f(D_{t_{i+1}})}{f\left(F^{-1}\left(x_i+F(D_{t_i})\right)\right)}\frac{\partial}{\partial
%% x_{i+1}}D_{t_n},\qquad i=0,\dots,n-2.
%% \end{equation}
%% Furthermore, according to~(\ref{costG}) and~(\ref{G
%% derivative}),
%% \begin{eqnarray}\label{dx_i C_0}
%% \lefteqn{\frac{\partial}{\partial
%% x_i}C^{(2)}=F^{-1}\left(x_i+F(D_{t_i})\right)+}\\
%% \nonumber&&\qquad \qquad +
%% \sum_{n=i+1}^N f(D_{t_n})\frac{\partial}{\partial x_i}D_{t_n} 
%%               \left[ F^{-1}\left(x_n+F(D_{t_n})\right)-D_{t_n} \right]
%% \end{eqnarray}
%% for~$i=0,\dots,N$.  Combining~(\ref{dx_i C_0}) and~(\ref{deltatrick}) yields~(\ref{lembeh}). Note that~(\ref{deltatrick}) is only valid up to $i=n-2$.
%% \hfill\sq
 
%% \medskip

\begin{lemma}\label{uniqueness_2}Under the assumptions of the
    Theorem~\ref{prop3}, equation~(\ref{x_0 condition 2}) has at most one
    solution~$x_0>0$. Besides, the function $g(x):= f(x)-{{a}} f({{a}} x)$ is positive.
\end{lemma}
 
\Proof 
Uniqueness will follow if we can show that both $h_2 \circ F^{-1}$ and 
\[\hat{h}_2(y):=-F^{-1}\left(X_0-N\left[y-F\left({{a}} F^{-1}(y)\right)\right]\right) \]
    are strictly increasing. Moreover,  $h_2 \circ F^{-1}(0)=0$ and
$\hat{h}_2(0)<0$ so that any zero of $h_2\circ F^{-1}+\hat h_2$ must be strictly positive.

The
    function $h_2$ is one-to-one, has zero as fixed point,  and
satisfies (\ref{h2increasing}).
It is therefore strictly increasing, and since $F^{-1}$ is also  strictly increasing,
we get that
    $h_2 \circ F^{-1}$ is  strictly increasing. It remains to  show that $\hat{h}_2$ is
 strictly increasing. We have that
 \[\hat{h}'_2(y)=N  \frac{f\left(F^{-1}(y)\right)-{{a}} f\left ({{a}}
F^{-1}(y)\right)}{f\left(F^{-1}(y)\right) f\left(F^{-1}\left(X_0-N\left[y-F\left({{a}}
F^{-1}(y)\right)\right]\right)\right)},\]
is strictly positive, because, as we will show now, the numerator of this
term is positive. The numerator can be expressed as $g(F^{-1}(y))$ for  $g$ as in the assertion.
Hence, establishing strict positivity of $g$ will conclude the proof. 
 To prove this we also define~$g_2(x):=f(x)-{{a}}^2 f({{a}} x)$ so that
\[h_2(x)=x\frac{g_2(x)}{g(x)}.\]
Both functions $g$ and $g_2$ are continuous and  have the same sign for all $x \in \R$
due to the properties of $h_2$ explained at the beginning of this proof. Because of
$g(x)<g_2(x)$ for all $x \in \R$, we infer that there can be no change of signs, i.e.,
either
$g(x)>0$ and $g_2(x)>0$ for all $x \in \R$ or $g(x)<0$ and $g_2(x)<0$ everywhere. With
$g(0)=f(0)(1-{{a}})>0$ we obtain the positivity of $g$.
\hfill\sq

Theorem~\ref{prop3} will now follow by combining the following proposition with the
arguments explained in Section~\ref{ReductionSection}.

\begin{proposition} The function $C^{(2)}:\Xi\to\R$ has the strategy $\xi^{(2)}$
from Theorem~\ref{prop3} as its unique minimizer.
Moreover, the components of $\xi^{(2)}$ are all strictly positive.
\end{proposition}
 
\Proof  The structure of the proof is similar to
the one of Theorem~\ref{prop1} although the computations are different. Thanks to
Lemma~\ref{lemma4}, we know that there exists 
 an optimal strategy $\xi^*=(x_0^*,\dots, x^*_N)\in\Xi$. There also exists a
corresponding Lagrange multiplier~$\nu$ such that
$$\frac{\partial}{\partial x_i}C^{(2)}(x_0^*,\dots, x^*_N)=\nu,\qquad i=0,\dots,N.$$
From~(\ref{lembeh}), we get  
\[ \nu=h_2\left(F^{-1}\left(x^*_i+F\left(D_{t_i}\right)\right)\right),  \qquad
i=0,\dots,N-1.\]
Since $h_2$ is one-to one, this implies in particular that $x^*_i+F\left(D_{t_i}\right)$
does not depend on $i=0,\dots,N-1$. It follows from~\eqref{delta dynamic}
also $D_{t_i+}=F^{-1}(x^*_i+F\left(D_{t_i}\right))$ is constant in $i$, and so
\begin{equation}\label{cons}
D_{t_i+}=D_{t_0+}=F^{-1}(x^*_0)\qquad\text{and}\qquad D_{t_{i+1}}=aF^{-1}(x^*_0).
\end{equation}
 Hence, 
\begin{eqnarray}\nonumber
x^*_0&=&F\left(h_2^{-1}(\nu)\right),\\
\label{opst} x^*_i&=& x^*_0-F(D_{t_i})= x^*_0-F\left({{a}} F^{-1}(x^*_0)\right) 
\hspace{0.5 cm} \mbox{for
$i=1,\dots,N-1$,}\\
\nonumber x^*_N&=&X^*_0-x^*_0-(N-1)\left[x^*_0-F\left({{a}} F^{-1}(x^*_0)\right)\right].
\end{eqnarray}
These equations link all market orders to the initial trade~$x^*_0$. Using~\eqref{opst} and once again~(\ref{cons}), we find that  $C^{(2)}(x_0^*,\dots,x_N^*)$ is 
equal to
\begin{eqnarray*}
C_0^{(2)}(x_0^*)&:=&C^{(2)}\Big(x^*_0,x^*_0 -F({{a}} F^{-1}(x^*_0)),\dots,
X_0-N x^*_0+(N-1)F
 ({{a}} F^{-1}(x^*_0) )\Big)\\
&=& N\left[G(x^*_0)-\widetilde{F}\left({{a}} F^{-1}(x^*_0)\right)\right]+
G\left(X_0+N\left[F\left({{a}} F^{-1}(x^*_0)\right)-x^*_0\right ]\right).
                    \end{eqnarray*}
The initial trade $x_0^*$ must clearly be a local minimum of $C_0^{(2)}$ and thus
$\frac{\partial}{\partial y}C^{(2)}_0(x^*_0)=0$. Therefore, 
$$0=N\left[D_{0+}-{{a}}^2 D_{0+}\frac{f(D_{t_1})}{f(D_{0+})}+D_{t_N+}\Big({{a}}
\frac{f(D_{t_1})}{f(D_{0+})}-1\Big)\right],$$
which is equivalent to
\begin{equation}\label{C20}
D_{t_N+}=D_{0+}\frac{f(D_{0+})-{{a}}^2 f(D_{t_1})}{f(D_{0+})-{{a}}
f(D_{t_1})}.
\end{equation}
This is just equation~\eqref{x_0 condition 2}, which  has at most one solution,
due  to Lemma~\ref{uniqueness_2}. This concludes the proof of the existence and the
representation of the optimal strategy $\xi^{(2)}$ in Theorem~\ref{prop3}.
 
Finally, we need to show the strict positivity of the optimal strategy. Thanks to the
positivity of the optimal~$x^*_0$, we get 
\[x^*_i=x^*_0-F({{a}} F^{-1}(x^*_0))>0\]
for~$i=1,\dots,N-1$. So it only remains to show that~$x^*_N>0$. 
We infer from~\eqref{C20} and~\eqref{cons} that
$$ D_{t_N+}=D_{0+}\frac{f(D_{0+})-{{a}}^2 f({{a}}  D_{0+})}{f(D_{0+})-{{a}} f({{a}}
D_{0+})} =D_{0+}\bigg[1+\frac{af(aD_{0+})-a^2f(aD_{0+})}{f(D_{0+})-af(aD_{0+})}\bigg].
$$
The fraction on the right is strictly positive due to   Lemma~\ref{uniqueness_2}.
Hence,
$$D_{t_N+}>D_{0+}=\frac1aD_{t_N}>D_{t_N},
$$
which implies $x_N^*>0$.\hfill \sq
\vskip4pt plus2pt

\section{Optimal strategy for block-shaped LOB.}\label{App_OW} Here we prove the
results of Section~\ref{Sec4}.
 
\medskip

%% \noindent{\it Proof of Corollary \ref{cor5}:} For a  constant LOB shape~$f\equiv q$ we
%% have~$F(x)=q x$ and~$F^{-1}(y)=\frac{y}{q}$. Since the corresponding functions~$h_1$
%% and~$h_2$ are one-to-one and~$\widetilde{F}(x)=\frac{q}{2}x^2$ is convex, we can apply
%% either Theorem~\ref{prop1} or Theorem~\ref{prop3}. Hence, the optimal initial
%% trade~$\xi^*_0$ can be computed by solving the equations~(\ref{x_0 condition 1})
%% or~(\ref{x_0 condition 2}). As explained in the remarks  following Theorems~\ref{prop1}
%% and~\ref{prop3}, the left-hand sides of these equations are equal to~$D_{t_N+}$. We have
%% \begin{eqnarray}
%% D_{0+}&=&\frac{\xi^*_0}{q},  \nonumber\\  D_{t_1}&=&{{a}} \frac{\xi^*_0}{q},\label{D
%% const}  \\
%% D_{t_N+}&=&{{a}} \frac{\xi^*_0}{q}+\frac{X_0-\xi^*_0-(N-1)\xi^*_0\left(1-{{a}} \right)}{q}
%% .\nonumber
%% \end{eqnarray}
%%  Equating the left-hand side of~(\ref{x_0 condition 1})
%% or~(\ref{x_0 condition 2}) with~$D_{t_N+}$ of~(\ref{D const}), we get
%% \[\xi^*_0=\frac{X_0}{(N-1)\left(1-{{a}}\right)+2}.\]
%% This is equivalent to
%% \begin{equation}\label{aux}
%% X_0=\xi^*_0 \left[ (N-1)\left(1-{{a}}\right)+2 \right].
%% \end{equation}
%% Applying Theorem~\ref{prop1} and~\ref{prop3}, we know that $\xi^*_1=\dots=\xi^*_{N-1}=\xi^*_0(1-{{a}})$. With this fact and~(\ref{aux}) we can compute
%% \[ \xi^*_N=X_0-(N-1)(1-{{a}})\xi^*_0-\xi^*_0=\xi^*_0. \]
%% \hfill\sq

%% \medskip

Our aim is to prove Proposition~\ref{propOW}, i.e., to show that the
strategy~(\ref{explicit strategy}) satisfies the recursion~(\ref{recursive
  strategy}). The key point is that we have indeed explicit formulas for the
coefficients in the backward schemes of Proposition~\ref{propOW}. 

\begin{lemma}\label{lemma8}{
The coefficients $\alpha_n$, $\beta_n$, and $\gamma_n$ from~(\ref{recursion1}) are
explicitly given by
\begin{eqnarray}\label{abc}
\alpha_n&=& \frac{\left(1+{{a}}^{-1}\right)-q \lambda \left[(N-n)\left({{a}}^{-1}-1\right)+2\left(1+{{a}}^{-1}\right)\right]}{2q\left[(N-n)\left({{a}}^{-1}-1\right)+\left(1+{{a}}^{-1}\right)\right]} \\
\nonumber \beta_n&=& \frac{1+{{a}}^{-1}}{\left[(N-n)\left({{a}}^{-1}-1\right)+\left(1+{{a}}^{-1}\right)\right]} \\
\nonumber \gamma_n&=& \frac{(N-n)\left(1-{{a}}^{-1}\right)}{2 \kappa \left[(N-n)\left({{a}}^{-1}-1\right)+\left(1+{{a}}^{-1}\right)\right]}. 
\end{eqnarray}
The explicit form of the sequences $\delta_n$, $\epsilon_n$ and $\phi_n$
from~(\ref{recursion2}) is
\begin{eqnarray}\label{def}
\delta_n&=& \frac{2{{a}}^{-2}\left[(N-n)\left({{a}}^{-1}-1\right)+\left(1+{{a}}^{-1}\right)\right]}{\kappa 
\left[(N-n)\left(1-{{a}}^{-2}\right)+(N-n+2)\left(a^{-3}-{{a}}^{-1}\right)\right]} \\
\nonumber \epsilon_n&=& \frac{\kappa\left({{a}}^{-1}-{{a}}\right)}{\left[(N-n)\left({{a}}^{-1}-1\right)+\left(1+{{a}}^{-1}\right)\right]} \\
    \nonumber \phi_n&=&
\frac{(N-n+1)\left({{a}}^{-1}-{{a}}\right)-(N-n)\left(1-a^2\right)}{ 
    \left[(N-n)\left({{a}}^{-1}-1\right)+\left(1+{{a}}^{-1}\right)\right]}.
\end{eqnarray}
}
\end{lemma}

{This Lemma can be proved in two steps. First, by a backward induction, we get the
explicit formulas for~$\alpha$, $\beta$ and~$\gamma$. Then, combining~\eqref{abc}
with~\eqref{recursion1} and~\eqref{recursion2}, we get~\eqref{def}. }

\medskip
 
\goodbreak\noindent{\it Proof of Proposition~\ref{propOW}.} We can deduce
 the following formulas from the preceding lemma:
\begin{equation}\label{intermed}
  \delta_n \epsilon_n=\frac{2}{(N-n)(1-a)+2}, \qquad \delta_n
\phi_n=\frac{2}{\kappa}\frac{(N-n)(1-a)+1}{(N-n)(1-a)+2}.
\end{equation}
They will turn out to be  convenient in~\eqref{recursive  strategy}. 
 
Let us
now consider the optimal strategy $(\xi^*_0,\dots,\xi^*_N)$ from~\eqref{explicit
strategy}. We consider the associated processes~$D_t:=D^A_t$ and
$X_t$ as defined in~\eqref{permanentimpactdynamics} and~\eqref{def_Xt}. For~$n=0$, we
have 
$$\xi^*_0=\frac{X_0}{(N-1)(1-a)+2}=\frac{1}{2}\delta_1 \epsilon_1$$ and it
satisfies~\eqref{recursive  strategy} because $D_0=0$. For $n\ge 1$, we can show
easily by induction on~$n$ that $D_{t_n}=a \kappa \xi^*_0$. From~\eqref{explicit
  strategy}, we get that~$\xi^*_n=(1-a)\xi^*_0$ for $n \not \in \{0,N\}$, and
therefore we get 
$$X_{t_n}=X_0-\xi^*_0-(n-1)(1-a)\xi^*_0=[(N-n)(1-a)+1]\xi^*_0.$$
 Using
these formulas, and combining with~\eqref{intermed}, it is now easy to check that\\
for $ n \in \{1,\dots,N-1\}$,
$$ \xi^*_n= \frac{1}{2}\left[
\delta_{n+1}\epsilon_{n+1}X_{t_n}-\delta_{n+1}\phi_{n+1}D_{t_n}\right],$$
which shows that the optimal strategy given in~\eqref{explicit strategy} solves~\eqref{recursive  strategy}.\sq 
\vskip4pt plus2pt
 
\end{appendix}
 
\medskip
 
\noindent{\bf Acknowledgement.}{ Support from the Deutsche Bank Quantitative
Products Laboratory is gratefully acknowledged.  The authors thank the
Quantitative Products Group of Deutsche Bank,  in particular Marcus Overhaus,
 Hans B\"{u}hler, Andy Ferraris, Alexander Gerko,  and Chrif Youssfi for stimulating discussions and useful comments
(the statements in this paper, however, express the private opinion of the authors and
do not necessarily reflect the views of Deutsche Bank). Moreover, it is a pleasure to thank
Anna Obizhaeva and Torsten
Sch\"oneborn for helpful comments on earlier versions of this paper.}

\end{document}